\documentclass[10pt, twocolumn, superscriptaddress, aps, pra, floatfix]{revtex4-1}

\usepackage{import}
\usepackage{preamble}
\usepackage{hyperref,url}

\makeatletter
\let\ORIbbl@fixname\bbl@fixname
\def\bbl@fixname#1{%
  \@ifundefined{languagealias@\expandafter\string#1}
    {\ORIbbl@fixname#1}
    {\edef\languagename{\@nameuse{languagealias@#1}}}%
}
\newcommand{\definelanguagealias}[2]{%
  \@namedef{languagealias@#1}{#2}%
}
\makeatother

\definelanguagealias{en}{english}

\def\corrAuthor{Angus J. Dunnett}

\def\corrEmail{angus.dunnett@insp.upmc.fr}

\begin{document}

\title{Simulating quantum vibronic dynamics at finite temperatures with many body wave functions at 0K} 

\author{Angus J. Dunnett}
\email{angus.dunnett@insp.upmc.fr}
\author{Alex W. Chin}
\affiliation{Sorbonne Universit\'{e}, CNRS, Institut des NanoSciences de Paris, 4 place Jussieu, 75005 Paris, France}

\begin{abstract}
  
  For complex molecules, nuclear degrees of freedom can act as an environment for the electronic `system' variables, allowing the theory and concepts of open quantum systems to be applied. However, when molecular system-environment interactions are non-perturbative and non-Markovian, numerical simulations of the complete system-environment wave function become necessary. These many body dynamics can be very expensive to simulate, and extracting finite-temperature results - which require running and averaging over many such simulations - becomes especially challenging. Here, we present numerical simulations that exploit a recent theoretical result that allows dissipative environmental effects at finite temperature to be extracted efficiently from a single, zero-temperature wave function simulation. Using numerically exact time-dependent variational matrix product states, we verify that this approach can be applied to vibronic tunneling systems and provide insight into the practical problems lurking behind the elegance of the theory, such as the rapidly growing numerical demands that can appear for high temperatures over the length of computations.
\end{abstract}
\maketitle

\section{Introduction}

\noindent

The dissipative quantum dynamics of electronic processes play a crucial role in the physics and chemistry of materials and biological life, particularly in the ultra-fast and non-equilibrium conditions typical of photophysics, nanoscale charge transfer and glassy, low-temperature phenomena \citep{miller_quantum_1983}. Indeed, the through-space tunneling of electrons, protons and their coupled dynamics critically determine how either ambient energy is transduced, or stored energy is utilised in supramolecular `devices', and real-time dynamics are especially important when the desired processes occur against thermodynamical driving forces, or at the single-to-few particle level \citep{devault1980quantum,may2008charge}. 

In many physio-chemical systems, a reaction, energy transfer, or similar event proceeds in the direction of a free
energy gradient, necessitating the dissipation of energy and the generation of entropy
\citep{dubi_colloquium_2011,benenti_fundamental_2017}. A powerful way of modelling the microscopic physics at work
during these irreversible dynamics is the concept of an `open' quantum system
\citep{breuer2002theory,weiss_quantum_2012}. Here a few essential and quantized degrees of freedom constituting the
`system' are identified and explicitly coupled to a much larger number of `environmental' degrees of freedom. Equations
of motion for the coupled system and environment variables are then derived and solved, with the goal of obtaining the
behaviour of the `system' degrees of freedom once the unmeasureable environmental variables are averaged over their
uncertain initial and final states. It is in this `tracing out' of the environment that the originally conservative,
reversible dynamics of the global system gives way to apparently irreversible dynamics in the behaviour of the system's
observable variables. The effective behaviour of the system `opened' to the environment is entirely contained within its
so-called reduced density matrix, which we shall later define. Important examples of the emergent phenomenology of
reduced density matrices include the ubiquitous processes of thermalization, dephasing and decoherence.
 
In the solid state, a typical electronic excitation will interact weakly with the lattice vibrations of the material, particularly the long-wavelength, low frequency modes. Under such conditions it is often possible to treat the environment with low-order perturbation theory and - given that the lattice `environment' relaxes back to equilibrium very rapidly - it is possible to derive a Markovian master equation for the reduced density matrix, such as the commonly used Bloch-Redfield theory \citep{breuer2002theory,may2008charge,weiss_quantum_2012}. However, in sufficiently complex molecular systems, such as organic bio-molecules, the primary environmental degrees of freedom acting on electronic states are typically the stochastic vibrational motions of the atomic nuclear coordinates. Unlike the solid state, these vibrations can: (1) couple non-perturbatively to electronic states, (2) relax back to equilibrium on timescales that are longer than the dynamics they induce in the system, and (3) have frequencies $\omega$ such that $\hbar\omega\gg K_B T$, where $T$ is the environmental temperature, and so must be treated quantum mechanically (zero-point energy and nuclear quantum effects). In this regime, the theory and numerical simulation of open quantum systems becomes especially challenging, as the detailed dynamics of the interacting system and environmental quantum states need to be obtained, essentially requiring the solution of a correlated (entangled) many body problem.  

One well known and powerful approach to this problem in theoretical chemistry is the Multi-layer Multiconfigurational Time-dependent Hartree (ML-MCTDH) technique, which enables vibronic wave functions to be efficiently represented and propagated without the \textit{a priori} limitations due to the `curse of dimensionality' associated with many body quantum systems \citep{wang2019quantum,lubich_time_2015-1}. However, computationally demanding methods based on the propagation of a large wave function from a definite initial state will typically struggle when dealing with finite-temperature environments (\textit{vide infra}), as the probability distribution of initial states requires extensive sampling. For this reason, the majority of ML-MCTDH studies have been effectively on zero-temperature systems. 

In this article we will explore a recent and intriguing development in an alternative approach to real-time dynamics and chemical rate prediction. This approach is based on the highly efficient representation and manipulation of large, weakly entangled wave functions with DMRG, Matrix-Product and Tensor-Network-State methods \citep{orus_practical_2014}. These methods, widely used in condensed matter, quantum information and cold atom physics, have recently been applied to a range of open system models, including chemical systems, but - as wave function methods - are typically used at zero-temperature \citep{prior2010efficient,prior_quantum_2013, chin2013role,xie_time-dependent_2019,alvertis_non-equilibrium_2019,schroder_tensor_2019}. However, a remarkable new result due to Tamascelli \textit{et al}. shows that it is indeed possible to obtain the \emph{finite-temperature} reduced dynamics of a system based on a simulation of a `pure', i.e. zero-temperature wave function \citep{tamascelli_efficient_2019}. 

In principle, this opens the way for many existing wave function methods to be extended into finite temperature regimes, although the present formulation of Tamascelli \textit{et al}.'s T-TEDOPA mapping is most easily implemented with matrix product states (MPS). In this article, we shall investigate this extension to finite temperature in the regime of relevance for molecular quantum dynamics, that is, non-perturbative vibrational environments, and present numerical data that verifies the elegance and utility of the method, as well as some of the potential issues arising in implementation. 

The structure of the article is as follows. In section \ref{sec:t-tedopa} we will summarise Tamascelli \textit{et al}.'s  T-TEDOPA mapping. In section \ref{sec:numerical-results} we verify the theory by comparing numerical simulations against an exactly solvable open system model, and also employ further numerical investigations to provide some insight into the manner in which finite temperatures are handled within this
method. By looking at the observables of the environment, we find that the number of excitations in the simulations grows continuously over time, which may place high demands on computational resources in some problems. In section \ref{sec:reaction-rates} we will present
results for a model system inspired by electron transfer in a multi-dimensional vibrational environment, and show how the temperature-driven transition from quantum tunneling to classical barrier transfer are successfully captured by this new approach. This opens a potentially fruitful new phase for the application of tensor network and related many body approaches for the simulation of non-equilibrium dynamics in a wide variety of vibronic materials and molecular reactions.

\section{T-TEDOPA}
\label{sec:t-tedopa}

\begin{figure}
  \includegraphics[width=\columnwidth]{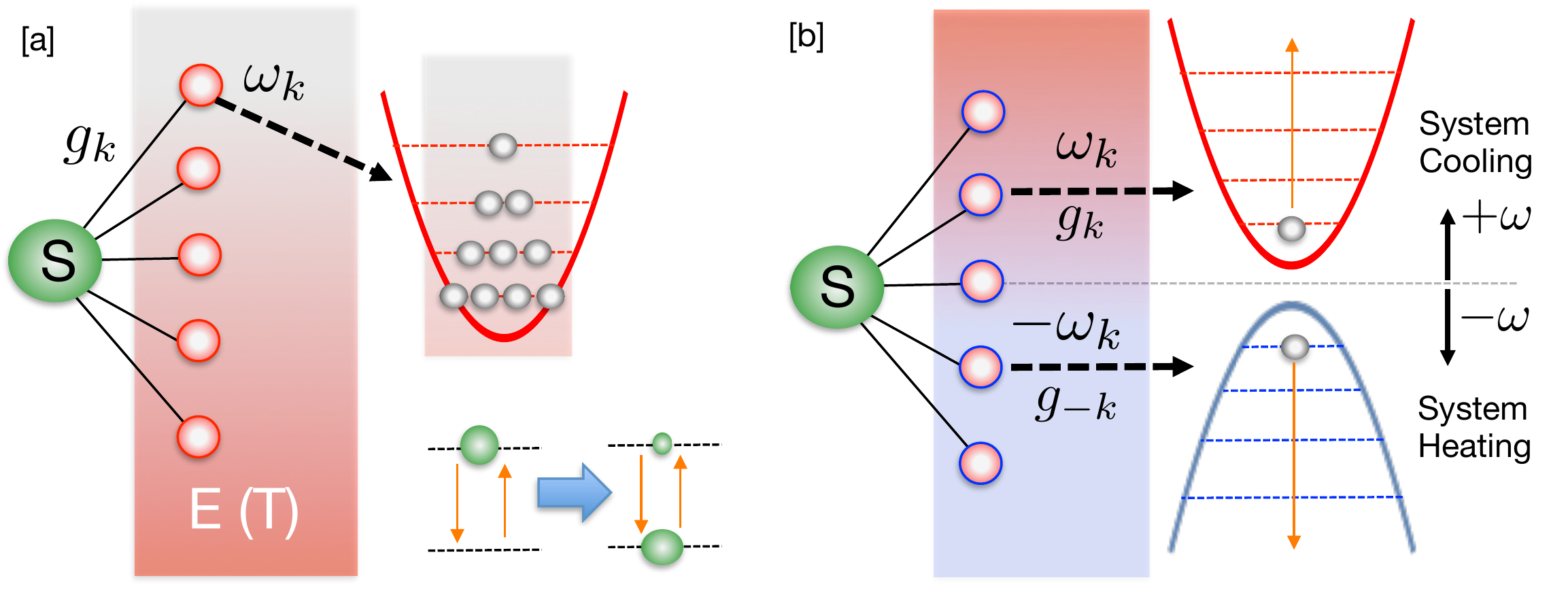}
  \caption{(a) A generic open quantum system contains a few-level `system' (S) that interacts with a much larger thermal heat bath of bosonic oscillators (the environment, E). The continuum of oscillator modes are initially uncorrelated with the system and each is thermally occupied with characteristic temperature $T=\beta^{-1}$. Coupling and stochastic fluctuations of the environment lead to the effective thermalization of the system, once the environmental states have been traced over. (b) In the T-TEDOPA approach, the harmonic environment is extended to include modes of negative frequency, and all modes  (positive and negative frequency) are initially in their ground states. It can be formally demonstrated that the thermalization of S in (a) can always be obtained from the pure zero-temperature state in (b), provided the spectral density of the original environment is known.}    
  \label{fig:tedopa1}
\end{figure}

\begin{figure}
  \includegraphics[width=\columnwidth]{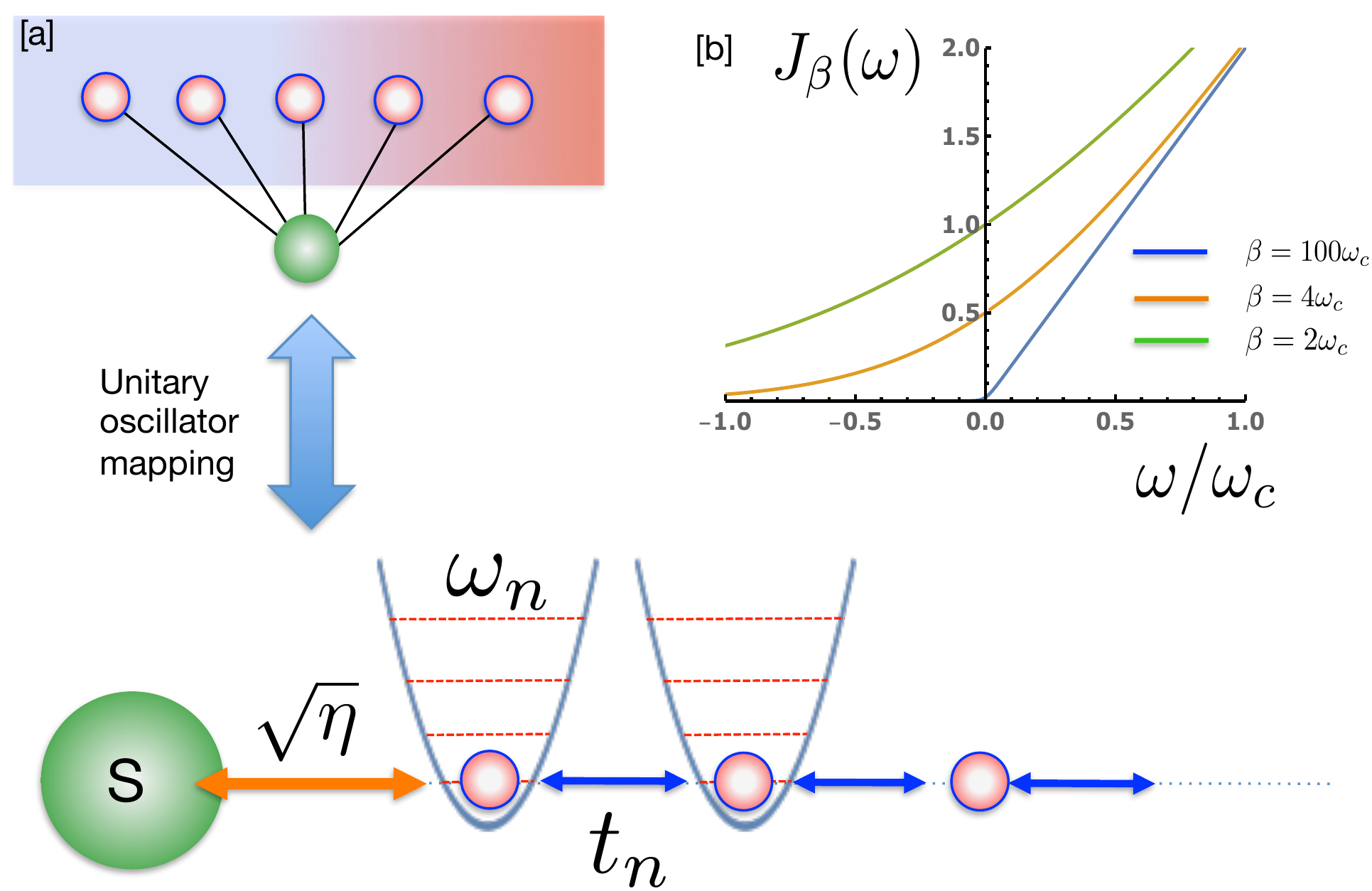}
  \caption{(a) The extended proxy environment of Fig. \ref{fig:tedopa1} (a) is described by an effective, temperature-dependent spectral density $J_\beta(\omega)$. Once the effective $J_\beta(\omega)$ has been specified, new oscillator modes can be found that provide a unitary transformation to a linear chain representation of the environment with nearest neighbour interactions. The non-perturbative wave function dynamics for such a many-body 1D system can be very efficiently simulated with MPS methods. (b) $J_\beta(\omega)$ for a physical Ohmic environment at three representative temperatures. At very low temperature ($\omega_c\beta\gg1$) there is essentially no coupling to the negative frequency modes, as excitation of these modes leads to an effective absorption of heat \emph{from} the environment. At higher temperatures, $J_\beta(\omega)$ becomes increasingly symmetric for the positive and negative modes.}
  \label{fig:tedopa2}
\end{figure}

\noindent
In this section we shall summarise the essential features of the T-TEDOPA approach, closely following the original notation and presentation of Tamascelli \textit{et al} \citep{tamascelli_efficient_2019}. Our starting point is the generic Hamiltonian for a system coupled to a bosonic environment consisting of
a continuum of harmonic oscillators
\begin{equation}
  \label{eq:ham}
  H_{SE}=H_{S}+H_{E}+H_{I},
\end{equation}
where
\begin{equation}
  \label{eq:Hi}
  H_{I}=A_{S}\otimes\int_{0}^{\infty}d\omega \hat{O}_{\omega},  H_{E} = \int_{0}^{\infty}d\omega \omega a_{\omega}^{\dagger}a_{\omega}.
\end{equation}
The Hamiltonian $H_{S}$ is the free system Hamiltonian, which for chemical systems, molecular photophysics and related problems will often be a description of a few of the most relevant diabatic states at some reference geometry of the environment(s) \citep{may2008charge}. $A_{S}$ is the system operator which couples to the bath. For the bath operators we take the displacements
\begin{equation}
  \label{eq:O}
  O_{\omega}=\sqrt{J(\omega)}(a_{\omega}+a_{\omega}^{\dagger}),
\end{equation}
thus defining the spectral density $J(\omega)$. This has been written here as a continuous function, but coupling to a discrete set of vibrational modes in, say, a molecular chromophore, can be included within this description by adding suitable structure to the spectral density, i.e. sets of Lorentzian peaks or Dirac functions \citep{wilhelm2004spin,schulze2015explicit,mendive2018multidimensional}. The state of the system+environment at time $t$ is described by a mixed state described by a density matrix
$\rho_{SE}(t)$. The initial condition is assumed to be a product of system and environment states
$\rho_{SE}(0)=\rho_{S}(0)\otimes\rho_{E}(0)$ where $\rho_{S}(0)$ is an arbitrary density matrix for the system and
$\rho_{E}(0) = \exp(-H_{E}\beta)/\mathcal{Z}$, with the environment partition function given by
$\mathcal{Z}=\Tr\{\exp(-H_{E}\beta)\}$. Such a product state is commonly realised in photophysics, where the reference geometry for the environment is the electronic ground state and the electronic system is excited according to the Franck-Condon principle into some manifold of electronic excited states without nuclear motion \citep{may2008charge,mukamel1995principles}. Indeed, this can also occur following any sufficiently rapid non-adiabatic event, just as ultra-fast charge separation at a donor-acceptor interface \citep{gelinas2014ultra-fast,smith2015phonon}. The environment thus begins in a thermal equilibrium state with inverse temperature $\beta$, and the energy levels of each harmonic mode are statistically populated, as shown in Fig. \ref{fig:tedopa1}a. For a very large (continuum) of modes, the number of possible thermal configurations of the initial probability distribution grows extremely rapidly with temperature, essentially making a naive sampling of these configurations impossible for full wave function simulations. We note, however, that some significantly better sampling methods involving sparse grids and/or stochastic mean-field approaches have been proposed and demonstrated \citep{alvermann2009sparse,binder2019first}. 

The initial thermal condition of the environmental oscillators is also a Gaussian state, for which is it further known that the influence functional \citep{weiss_quantum_2012} - which is a full description of
the influence of the bath on the system - will depend only on the two-time correlation function of the bath operators
\begin{equation}
  \label{eq:S}
  S(t)=\int_{0}^{\infty}d\omega\langle O_{\omega}(t)O_{\omega}(0)\rangle.
\end{equation}
Any two environments with the same $S(t)$ will have the same influence functional and thus give rise to the same reduced
system dynamics, i.e. the same $\rho_{S}(t)=\Tr\{\rho_{SE}(t)\}$. That the reduced systems dynamics are completed specified by the spectral density and temperature of a Gaussian environment has been known for a long time \citep{weiss_quantum_2012}, but the key idea of the equivalence - and thus the possibility of the interchange - of environments with the same correlation functions has only recently been demonstrated by Tamascelli \textit{et al}. \citep{tamascelli2018nonperturbative}. 

The time dependence in \eq{S} refers to the interaction picture so that the bath operators evolve under the free bath
Hamiltonian: $O_{\omega}(t)=e^{iH_{E}t}O_{\omega}(0)e^{-iH_{E}t}$. Using \eq{O} and
$\langle a_{\omega}^{\dagger}a_{\omega}\rangle = n_{\beta}(\omega)$ we have
\begin{equation}
  \label{eq:S2}
  S(t)=\int_{0}^{\infty}J(\omega)[e^{-i\omega t}(1+n_{\beta}(\omega))+e^{i\omega t}n_{\beta}(\omega)].
\end{equation}
Making use of the relation
\begin{equation}
  \label{eq:coth}
  \frac{1}{2}(1+\coth(\omega\beta/2)) \equiv
  \begin{cases}
    n_{\omega}(\beta), \omega \ge 0 \\
    -(n_{|\omega|}(\beta)+1), \omega < 0
  \end{cases}
\end{equation}
we can write \eq{S2} as an integral over all positive and negative $\omega$
\begin{equation}
  \label{eq:S3}
  S(t)=\int_{-\infty}^{\infty}d\omega\sign(\omega)\frac{J(|\omega|)}{2}(1+\coth(\frac{\omega\beta}{2}))e^{-i\omega t}.
\end{equation}
But \eq{S3} is exactly the two-time correlation function one would get if the system was coupled to a bath, now containing
positive and negative frequencies, at zero temperature, with a temperature weighted spectral density given by
\begin{equation}
  \label{eq:Jtherm}
  J_{\beta}(\omega)=\sign(\omega)\frac{J(|\omega|)}{2}(1+\coth(\frac{\omega\beta}{2})).
\end{equation}
Thus, we find that our open system problem is completely equivalent to the one governed by the Hamiltonian
\begin{equation}
  \label{eq:Hext}
  H = H_{S}+H_{E}^{\text{ext}}+H_{I}^{\text{ext}},
\end{equation}
in which the system couples to an extended environment, where
\begin{equation}
  \begin{split}
    \label{eq:ext}
  &H_{I}^{\text{ext}}=A_{S}\otimes\int_{-\infty}^{\infty}d\omega\sqrt{J_{\beta}(\omega)}(a_{\omega}+a_{\omega}^{\dagger}),\\
  &H_{E}^{\text{ext}}=\int_{-\infty}^{\infty}d\omega\omega a_{\omega}^{\dagger}a_{\omega},
\end{split}
\end{equation}
and which has the initial condition $\rho_{SE}(0)=\rho_{S}(0)\otimes\ket{0}_{E}\bra{0}$. The system now couples to a bath
consisting of harmonic oscillators of positive and negative frequencies which are initially in their ground states, as
shown in Fig. \ref{fig:tedopa1}b. This transformed initial condition is now far more amenable to simulation as the environment is now
described by a \emph{pure}, single-configuration wave function, rather than a statistical mixed state, and so \emph{no}
statistical sampling is required to capture the effects of temperature on the reduced dynamics!

Analysing the effective spectral density of Eq. \ref{eq:Jtherm}, it can be seen that the new extended environment has thermal detailed balance between absorption and emission processes encoded in the ratio of
the coupling strengths to the positive and negative modes in the extended \emph{Hamiltonian}, as opposed to the operator statistics of a thermally occupied \emph{state} of the original, physical mode, i.e.

\begin{equation}
\frac{J_{\beta}(\omega)}{J_{\beta}(-\omega)}=\frac{\langle a_{\omega}a^{\dagger}_{\omega}\rangle_\beta}{\langle a_{\omega}^{\dagger}a_{\omega}\rangle_\beta}=e^{\beta\omega}
\end{equation}

Indeed, from the
system's point of view, there is no difference between the absorption from an occupied, positive energy, bath mode and
the emission into an unoccupied, negative energy, bath mode.

In fact, the equivalence between these two environments
goes beyond the reduced system dynamics as there exists a unitary transformation which links the extended environment to
the original thermal environment. This means that one is able to reverse the transformation and calculate thermal
expectations for the actual bosonic bath such as $\langle a_{\omega}^{\dagger}(t)a_{\omega}(t)\rangle_{\beta}$. This is particularly useful for molecular systems in which environmental (vibrational) dynamics are also important observables that report on the mechanisms and pathways of physio-chemical transformations \citep{musser2015evidence,schnedermann2016sub,schnedermann2019molecular}. This is a major advantage of many body wave function approaches, as the full information about the environment is available, c.f. effective master equation descriptions which are obtained \emph{after} averaging over the environmental state. We note that the idea of introducing a second environment of negative frequency oscillators to provide finite temperature effects in pure wave functions was previously proposed in the thermofield approach of De Vegas and Banulus \citep{de_vega_thermofield-based_2015}. This approach explicitly uses the properties of two-mode squeezed states to generate thermal reduced dynamics, but the original thermofield approach, unlike the T-TEDOPA mapping, considered the positive and negative frequency environments as two separate baths.

Following this transformation a further step is required to facilitate efficient simulation of the many-body
system+environment wave function. This is to apply a unitary transformation to the bath modes which converts the
\emph{star}-like geometry of $H_{I}^{\text{ext}}$ into a \emph{chain}-like geometry, thus allowing the use of MPS methods \citep{chin2010exact,chin2013role, prior_quantum_2013}. We thus define new modes
$c_{n}^{(\dagger)}=\int_{-\infty}^{\infty}U_{n}(\omega)a_{\omega}^{(\dagger)}$, known as chain modes, via the unitary
transformation $U_{n}(\omega)=\sqrt{J_{\beta}(\omega)}p_{n}(\omega)$ where $p_{n}(\omega)$ are orthonormal polynomials with
respect to the measure $d\omega J_{\beta}(\omega)$. Thanks to the three term recurrence relations associated with all
orthonormal polynomials $p_{n}(\omega)$, only one of these new modes, $n=1$, will be coupled to the system, while all
other chain modes will be coupled only to their nearest neighbours \citep{chin2010exact}. Our interaction and bath Hamiltonians thus become
\begin{equation}  
\begin{split}
  \label{eq:chain}
  & H_{I}^{\text{chain}}=\kappa A_{S}(c_{1}+c_{1}^{\dagger}),\\
  & H_{E}^{\text{chain}}=\sum_{n=1}^{\infty}\omega_{n}c_{n}^{\dagger}c_{n}+\sum_{n=1}^{\infty}(t_{n}c_{n}^{\dagger}c_{n+1}+h.c).
\end{split}
\end{equation}
The chain coefficients appearing in \eq{chain} are related to the three-term recurrence parameters of the orthonormal polynomials
and can be computed using standard numerical techniques \citep{chin2010exact}. The full derivation of the above Hamiltonian is given in the appendix. Since the initial state of the bath was the vacuum state, it is
unaffected by the chain transformation. 

We have thus arrived at a formulation of the problem of finite-temperature open systems in which the many-body environmental state is initialised as a pure product of trivial ground states, whilst the effects of thermal fluctuations and populations are encoded in the Hamiltonian chain parameters and system-chain coupling. These parameters must be determined once for each temperature but - in principle - the actual simulation of the many body dynamics is now no more complex than a zero-temperature simulations. This thus opens up the use of powerful $T=0K$ wave function methods for open systems, such as those based on MPS, numerical renormalisation group and ML-MCTDH \citep{wang2019quantum,lubich_time_2015-1}. However, while this seems remarkable - and we believe this mapping to be a major advance - there must be a price to be paid elsewhere. We shall now demonstrate with numerical examples where some of the computational costs for including finite-$T$ effects may appear and discuss how they might effect the feasibility and precision of simulations. We also propose a number of ways to mitigate these potential problems within the framework of tensor network approaches.

\section{Numerical tests and computational efficiency}
\label{sec:numerical-results}

All numerical results in the following sections are obtained by representing the many body system-environment wave function as a MPS and evolving it using time-dependent variational methods. All results have been converged w.r.t. the parameters of MPS wave functions (bond dimensions, local Hilbert space dimensions, integrator time steps), meaning that the results and discussion should - unless explicitly stated - pertain to the essential properties of the T-TEDOPA mapping itself. Extensive computational details and background theory can be founds in Refs. \citep{orus_practical_2014,schollwock_density-matrix_2011,lubich_time_2015,paeckel_time-evolution_2019,haegeman_unifying_2016}. 

\subsection{Chain dynamics and chain-length truncation}
\begin{figure}
  \includegraphics[width=\columnwidth]{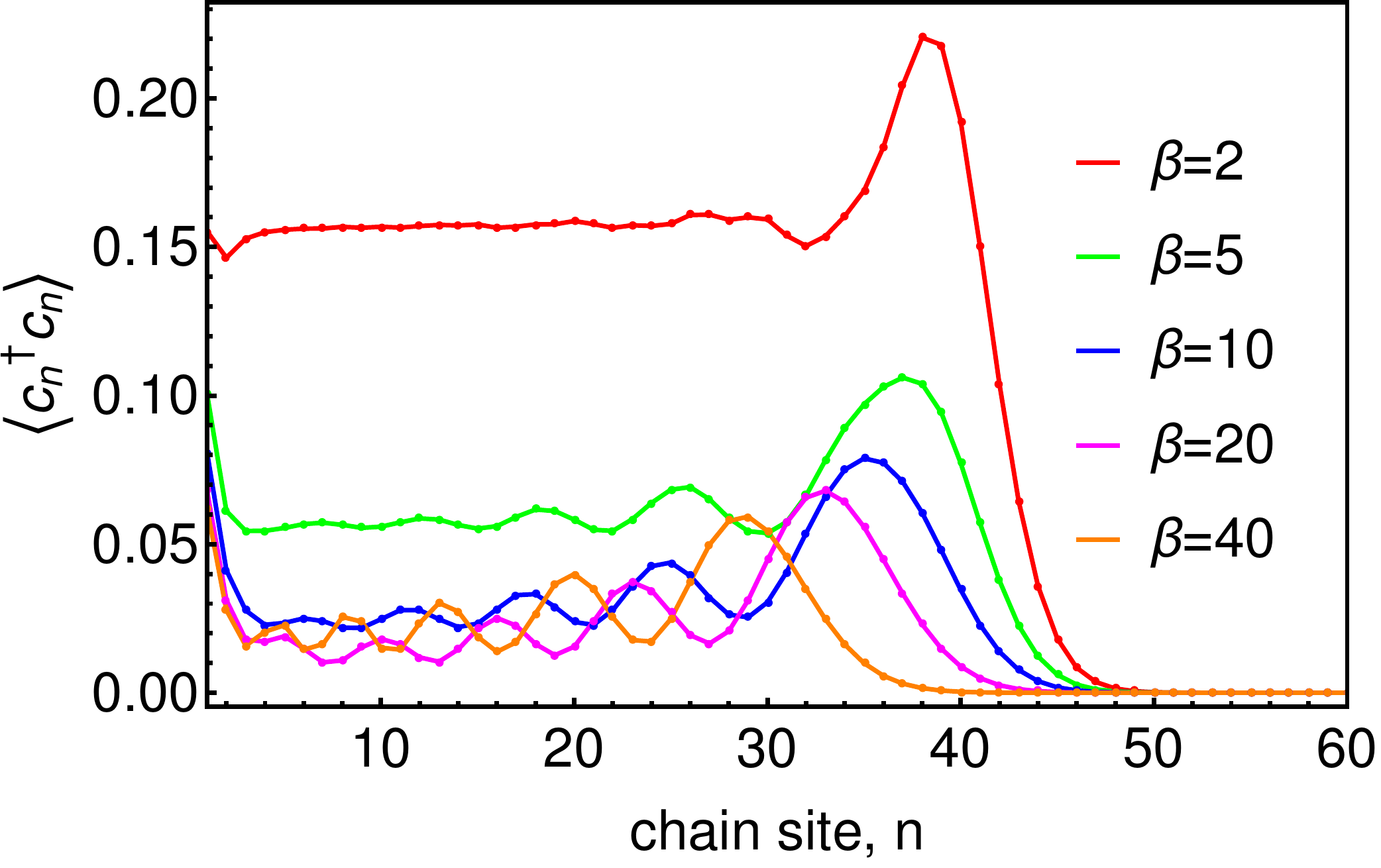}
  \caption{Chain mode occupations $\langle c_{n}^{\dagger}c_{n}\rangle$ at time $\omega_{c}t=45$ for baths of several
    temperatures. The system, which in this case is the Ohmic SBM, with $\omega_{0}=0.2\omega_{c}$ and $\alpha=0.1$, is
    attached at site $n=1$ of the chain.}
  \label{fig:chain}
\end{figure}

Before looking at the influence of thermal bath effects on a quantum system, we first investigate the effects of the changing chain parameters that appear due to the inclusion of temperature in the effective spectral density $J_\beta(\omega)$. As a consequence of the nearest-neighbour nature of \eq{chain} (see Fig. \ref{fig:tedopa2}), the chain mapping establishes a kind of causality among
the bath modes which is extremely convenient for simulation. Starting from $t=0$ the system will interact first with the
chain mode $n=1$ which, as well as acting back on the system, will in turn excite the next mode along the chain and so
on. The dynamics thus have a well defined light-cone structure in which a perturbation travels outwards from the system
along the chain to infinity. This means that we may truncate the chain at any distant mode $n=N$ without causing an error in
the system or bath observables up to a certain time $T_{LC}(N)$ which is the time it takes for the edge of the light-cone
to reach the $Nth$ chain mode. Beyond $T_{LC}(N)$ there will be reflections off the end of the chain leading to error in
the bath observables, however these reflections will not cause error in the system observables until the time
$t\approx 2T_{LC}(N)$. Figure \ref{fig:chain} shows a snapshot of the chain mode occupations for the Ohmic spin-boson model
considered in the next section. One can see that the velocity of the wave-front that travels outward from the system
depends on temperature, with hotter baths leading to faster propagation and thus requiring somewhat longer chains. 

To enable simulation we are also required to truncate the infinite Fock space dimension of each chain mode to a finite
dimension $d$, introducing an error for which there exist rigorously derived bounds \citep{woods_simulating_2015}. The
initial state $\ket{\Psi(0)}_{SE} = \ket{\psi(0)}_{S}\otimes \ket{0}_{E}$ (here we specialize to the case where the
system is initially in a pure state) can then be encoded in an MPS and evolved under one of the many time-evolution
methods for MPS. We choose to use the one-site Time-Dependent-Variational-Principle (1TDVP) as it has been shown to be a
efficient method for tracking long-time thermalization dynamics and has previously been shown to give numerically exact results for the zero-temperature spin-boson model in the highly challenging regime of quantum criticality \citep{schroder2016simulating}. In our implementation of
1TDVP the edge of the light-cone is automatically estimated throughout the simulation by calculating the overlap of the
wave-function $\ket{\Psi(t)}_{SE}$ with its initial value $\ket{\Psi(0)}_{SE}$ at each chain site. This allows us expand
the MPS dynamically to track to expanding light-cone, providing roughly a 2-fold speed-up compared to using a fixed
length MPS.

\subsection{Two-level system dynamics: dephasing and divergence of chain occupations due to energy exchange}
\begin{figure}
  \includegraphics[width=\columnwidth]{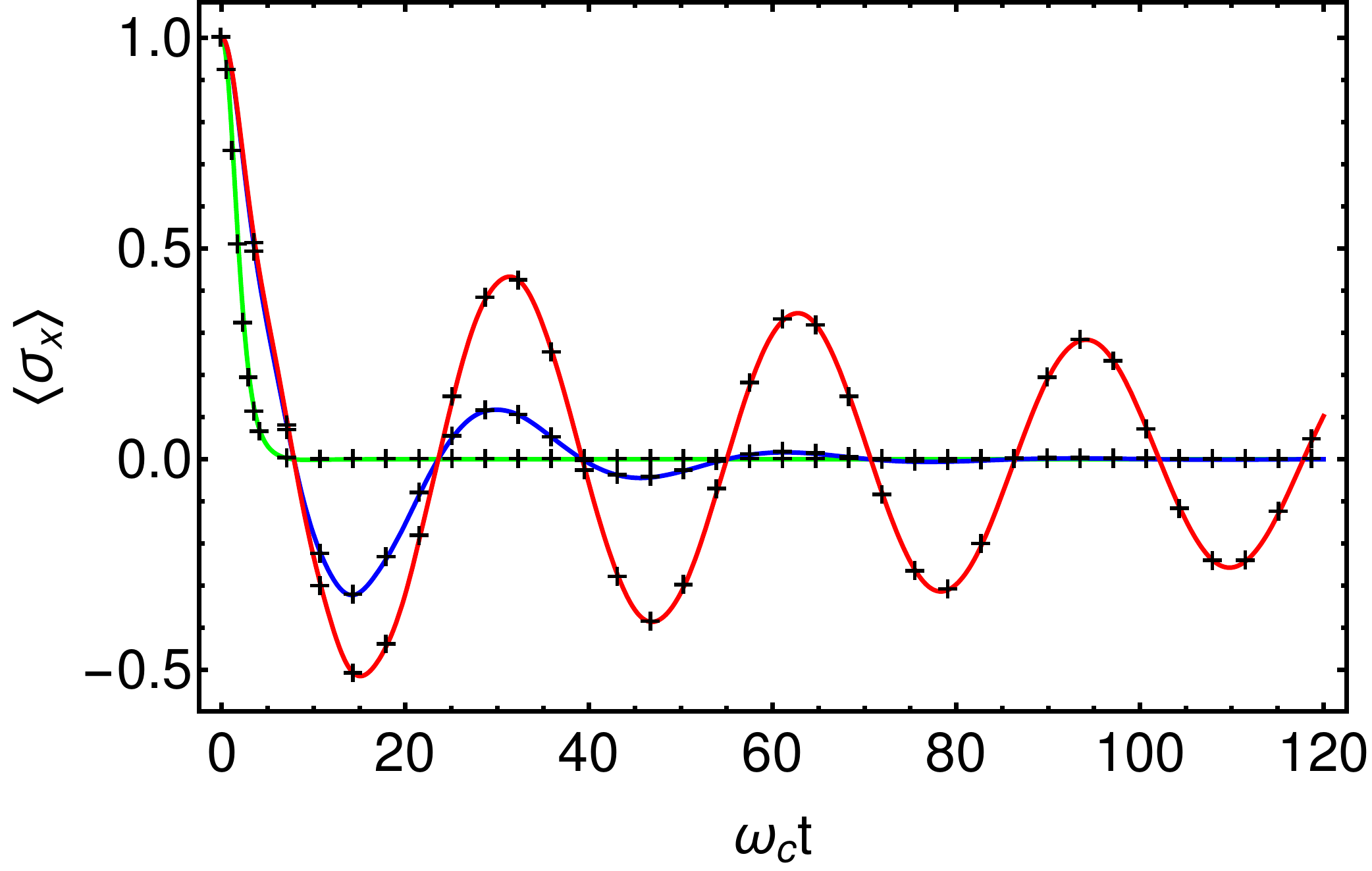}
  \caption{Comparison of T-TEDOPA (Black crosses) with the exact solution for the Independent-Boson-Model at $\beta=100$ (Red),
    $\beta=10$ (Blue) and $\beta=1$ (Green). $H_{S}=\frac{\omega_{0}}{2}\sigma_{z}$, $A_{S}=\sigma_{z}$,
    $J(\omega) = 2\alpha\omega_{c}(\frac{\omega}{\omega_{c}})^{s}\Theta(\omega-\omega_{c}))$, $\alpha=0.1$, $s=1$,
    $\omega_{0}=0.2\omega_{c}$}
  \label{fig:ibm}
\end{figure}

\noindent
To confirm the accuracy of this approach in terms of reduced system dynamics we now explore the effects of a dissipative environment on a quantum two-level system. First, we compare the numerical results against the analytically solvable Independent-Boson-Model (IBM) \citep{mahan_many-particle_2000,breuer2002theory}. This is a model of pure dephasing, defined
by $H_{S}=\frac{\omega_{0}}{2}\sigma_{z}$ and $A_{S}=\sigma_{z}$, where $\{\sigma_{x},\sigma_{y},\sigma_{z}\}$ are the standard Pauli matrices. We take an Ohmic spectral density with a hard cut-off
$J(\omega)=2\alpha\omega\Theta(\omega-\omega_{c})$ and choose a coupling strength of $\alpha=0.1$ and a gap of
$\omega_{0}=0.2\omega_{c}$ for the two level system (TLS). The initial state of the system is a positive superposition of the spin-up and spin-down states, and we monitor the decay of the TLS coherence, which is quantified by $\langle \sigma_x (t)\rangle$.
All results were converged using a Fock space dimension of $d=6$ for
the chain modes and maximum MPS bond-dimension $D_{\text{max}}=4$. We find that the results obtained using the T-TEDOPA
method agree very well with the exact solution (see \fig{ibm}) and correctly reproduce the transition from under-damped
to over-damped decay as the temperature is increased \citep{mahan_many-particle_2000,breuer2002theory}. 

\begin{figure}
  \includegraphics[width=\columnwidth]{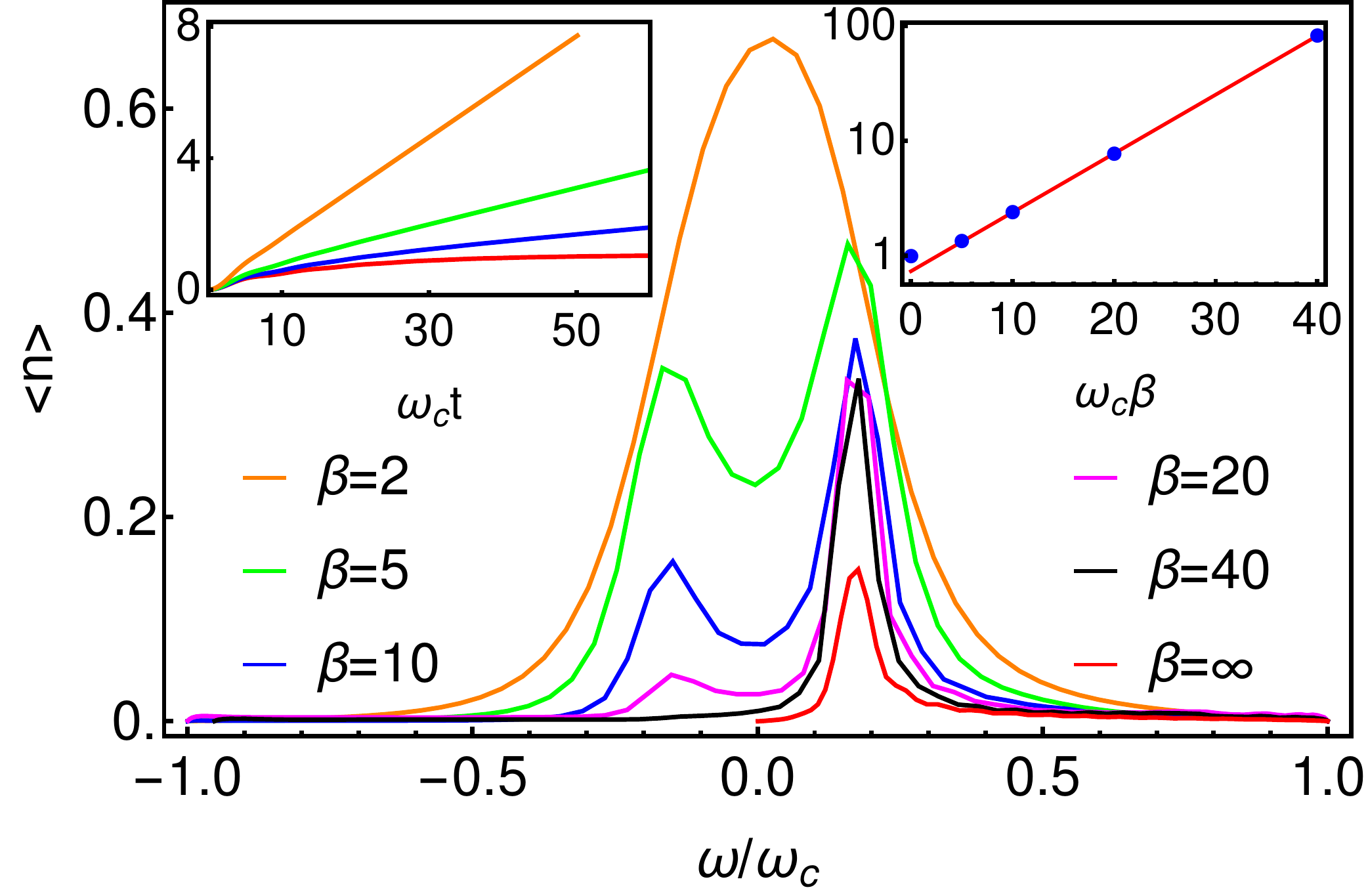}
  \caption{Bath mode occupations $\langle n_{\omega}\rangle=\langle a_{\omega}^{\dagger}a_{\omega}\rangle$ for the extended
    environment after the TLS has decayed. The TLS is governed by a Hamiltonian $H_{S}=\frac{\omega_{0}}{2}\sigma_{z}$
    where $\omega_{0}=0.2\omega_{c}$ and is coupled to an Ohmic bath with a hard cut-off via $A_{S}=\sigma_{x}$. The
    coupling strength is $\alpha=0.1$. Left inset: total mode occupation as a function of time
    $\langle n\rangle_{\text{tot}} = \int_{-\infty}^{\infty}d\omega\langle n_{\omega}\rangle$. Right inset shows
    $\frac{\langle n_{\omega_{p}}\rangle+1}{\langle n_{\omega_{n}}\rangle}$ plotted on a log scale against the inverse
    temperature, demonstrating the detailed balance of the absorption and emission rates.}
    \label{fig:db}
\end{figure}

As a second numerical example we take the Spin-Boson-Model (SBM), identical to the IBM considered above except that now
the TLS couples to the bath via $A_{S}=\sigma_{x}$. Unlike the previous case, the bath can now drive transitions within the TLS, so that energy is now dynamically exchanged between the TLS and its environment. Indeed, as $A_{S}$ no longer commutes with $H_S$, no exact solution for this model is known \citep{weiss_quantum_2012}. It has thus become an important testing ground for numerical approaches to non-perturbative simulations of open systems and has been widely applied to the physics of decoherence, energy relaxation and thermalization in diverse physical, chemical and biological systems - see Refs. \citep{weiss_quantum_2012,de2017dynamics} for extensive references. In our example, we prepare the spin in the upper spin state ($\langle \sigma_z \rangle=+1$) and allow the bath to thermalize by environmental energy exchange (see Fig. \ref{fig:tedopa1}a). Here, instead of presenting the spin dynamics for this model we will here
interest ourselves in the observables of the bath as these will
provided insight into the manner in which a finite temperature bath is being mimicked by an initially empty tight-binding
chain. In figure \ref{fig:db} we plot the bath mode occupations $\langle a_{\omega}^{\dagger}a_{\omega}\rangle$ for several
temperatures. Each observation was taken after the spin had decayed into its thermal steady state and thus provides a kind of
absorption spectrum for the system. We note that these data refer to the modes of the extended environment of \eq{Hext}
rather than the original bosonic bath and thus the mode energies run from $-\omega_{c}$ to $\omega_{c}$.

We find that for zero temperature ($\beta=\infty$) the bath absorption spectrum contains a single peak at a frequency
around $\omega_{p}=0.17\omega_{c}$, suggesting that the spin emits into the bath at a re-normalized frequency that is lower than the bare
gap of the TLS ($\omega_0=0.2\omega_c$). This agrees well with the renormalized gap
$\omega_{0}^{r}=\omega_{0}(\omega_{0}/\omega_{c})^{\frac{\alpha}{1-\alpha}}$ predicted by the non-perturbative variational polaron theory of Silby \& Harris
\citep{silbey_variational_1984}, which for the parameters used here gives $\omega_{0}^{r}=0.167\omega_{c}$.

Moving to non-zero temperature we see that a peak begins to form at a corresponding negative frequency, which we interpret as being due the spin absorbing thermal energy from the
bath by the \emph{emission} (creation) of negative energy quanta. In accordance with detailed balance, the ratio between the positive and negative frequency peaks approaches unity
as temperature is increased and by $\beta\omega_c =2$ the two peaks have merged to form a single, almost symmetric, distribution,
reflecting the dominance of thermal absorption and emission over spontaneous emission at high temperature. Indeed, as
shown in the right inset of figure \ref{fig:db} the ratio of the peak heights we extract obeys
$\frac{\langle n_{\omega}\rangle+1}{\langle n_{-\omega}\rangle} = e^{\epsilon\beta}$ with $\epsilon=0.118$. Thus we see
that the chain is composed of two independent vacuum reservoirs of positive and negative energy which the system emits into at
rates which effectively reproduce the emission and absorption dynamics that would be induced by a thermal bath.

However, the introduction of positive and negative modes has an interesting and important consequence for the computational
resources required for simulation. Shown in the left inset of figure \ref{fig:db} is the total mode occupation as a
function of time for some of the different temperatures simulated. One sees that for $\beta=\infty$  (zero temperature) the total occupation
of the bath modes increases initially and then plateaus at a certain steady state value corresponding to the total
number of excitations created in the bath by the TLS during its decay. In contrast, for finite temperature, the total mode occupation
increases indefinitely at a rate which grows with temperature. This is despite the fact that for the finite temperature
baths the total excitation number will also reach a steady state once the TLS has decayed. The reason for this is clear.
The thermal occupation of the physical bath mode with frequency $\omega$ is obtained by subtracting its negative,
from its positive energy counterpart in the extended mode basis, i.e.
$\langle n_{\omega}\rangle_{\beta}=\langle n_{\omega}\rangle_{\ket{0}_{E}}-\langle n_{-\omega}\rangle_{\ket{0}_{E}}$. While
$\langle n_{\omega}\rangle_{\beta}$ will reach a steady state, the components $\langle n_{\omega}\rangle_{\ket{0}_{E}}$ and
$\langle n_{-\omega}\rangle_{\ket{0}_{E}}$ will be forever increasing, reflecting the fact that the TLS reaches a
\emph{dynamic} equilibrium with the bath in which energy is continuously being absorbed from and emitted into the bath
at equal rates, thus filling up the positive and negative reservoirs. Since it is the modes of the \emph{extended}
environment that appear in the numerical simulation, one will always encounter potentially large errors once the filling of the modes
exceeds their capacity set by the truncation to $d$ Fock states per oscillator. The rate at which this filling occurs increases with temperature and is linear in time. However, as the relaxation time of the system is also broadly proportional to temperature for $\beta\omega_{c}\ll1$, this may not be a problem, if one is only interested in the short-time transient dynamics. Where this may pose problems is for the extraction of converged properties of relaxed, i.e. locally thermalized excited states, such as their (resonance) fluorescence spectra, or multidimensional optical spectra \citep{mukamel1995principles}. While these ever-growing computational resources must - as argued above - be present in any simulation approach, we note that one possible way to combat the growth of local dimensions could be to use the dynamical version of Guo's Optimised Boson Basis (OBB) which was introduced into 1TDVP for open systems by Schroeder \textit{et al.} \citep{guo2012critical,schroder2016simulating}.   

\section{Electron Transfer}
\label{sec:reaction-rates}
\begin{figure}
  \includegraphics[width=\columnwidth]{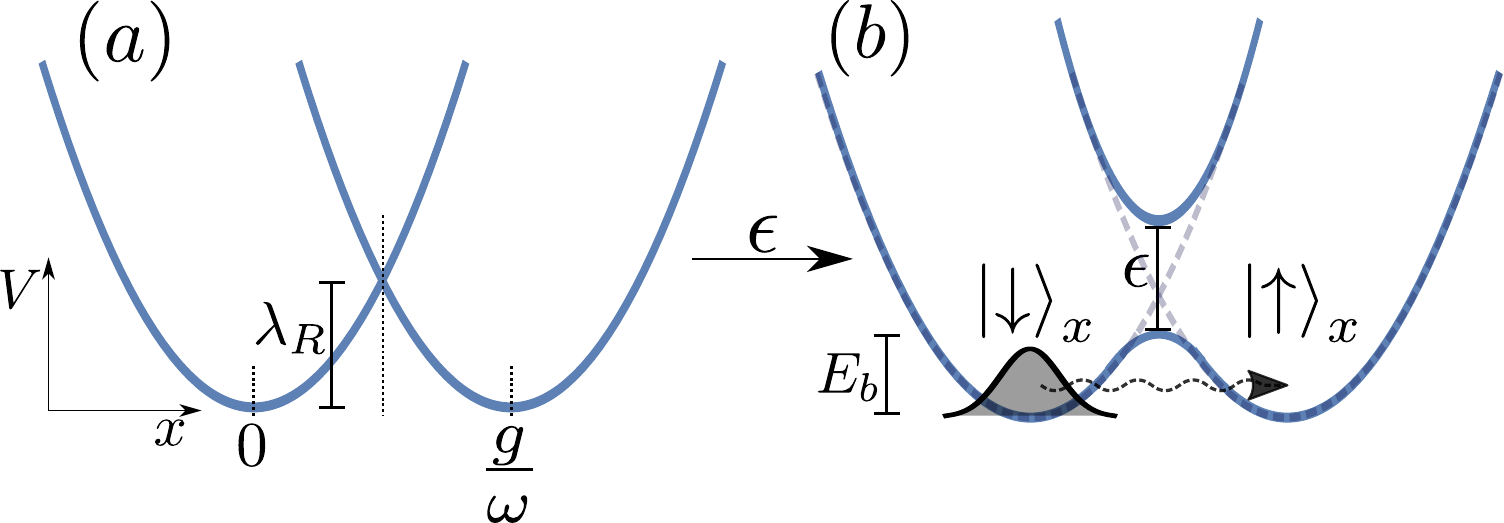}
  \caption{(a) Potential energy surfaces (Marcus parabolas) for $\epsilon=0$ as a function of the reaction coordinate
    $x$. We consider only the case of zero bias, i.e. when the minima of the two wells are at the same energy. (b) Turning
    the electronic coupling $\epsilon$ leads to an avoided crossing and thus an energy barrier $E_{b}$ for the reaction.
    Note that this is a simplified picture in which we treat the bath as being represented by a single mode of frequency
    $\omega$ and coupling strength $g$ whereas in the actual model we simulate there is a similar surface for all bath
    modes.}
  \label{fig:schem}
\end{figure}

\noindent
Having established that the T-TEDOPA mapping allows efficient computational access to finite temperature open dynamics, we now study the chemically relevant problem of tunneling electron transfer. Electron transfer is a fundamental
problem in chemical dynamics and plays an essential role in a vast variety of crucial processes including the
ultra-fast primary electron transfer step in photosynthetic reaction centers and the electron transport that powers biological respiration  \citep{devault1980quantum,marcus1993electron, may2008charge}. The problem of modeling electron transfer between molecules
comes down to accurately treating the coupling between the electronic states and environmental vibrational modes, and often involves the use of first principle techniques to parameterize the total spectral functions of the vibrational and outer solvent, or protein environment \citep{mendive2018multidimensional,schroder_tensor_2019,zuehlsdorff_optical_2019}. In many
molecular systems - and particularly biological systems where the transfer between electronic states is affected by coupling to chromophore and protein
modes - the system-bath physics is highly non-perturbative and $J(\omega)$ has very sharp frequency-dependence \citep{may2008charge,womick2011exciton,chin2013role,kolli2012fundamental}. Until recently, and even at zero temperature, a fully quantum mechanical description of the
coupling to a continuum of environmental vibrations was challenging due to the exponential scaling of the vibronic wave functions. However, with advances in numerical approaches driven by developments in Tensor-Networks and ML-MCTDH, the exact quantum simulation of continuum environment models can now be explored very precisely at zero temperature. Given this, we now explore how the T-TEDOPA mapping can extend this capability to finite temperature quantum tunneling.

Here, we will again adapt the spin-boson model to analyse a typical donor-acceptor electron transfer system, as shown in Fig.\ref{fig:schem}. In this model the electron transfer process is modelled
using two states representing the reactant and product states which we take to be the eigenstates of $\sigma_{x}$ with
$\ket{\downarrow}$ representing the reactant and $\ket{\uparrow}$ the product. We take our system Hamiltonian to be
$H_{S}=\frac{\epsilon}{2}\sigma_{z} + \lambda_{R}\frac{1+\sigma_{x}}{2}$, and the coupling operator as
$A_{S}=\frac{1+\sigma_{x}}{2}$, where $\lambda_{R}$ is the reorganization energy which for an Ohmic bath is
$\lambda_{R}=2\alpha\omega_{c}$. The electron tunnels from the \emph{environmentally relaxed} reactant state to the product state by moving through a
multi-dimensional potential energy landscape along a collective reaction coordinate which is composed of the displacements of the
ensemble of bath modes (this is effectively the coordinate associated with the mode that is directly coupled to the system in the chain representation of the environment). Figure \ref{fig:schem}(a) shows two potential energy surfaces - Marcus parabolas - of
the electronic system for $\epsilon=0$. Although in the actual model we simulate the reaction coordinate is composed of
the displacements of an infinite number of modes, in figure \ref{fig:schem} we present a simplified picture in which the
electron moves along a single reaction coordinate, $x$. The potential minimum of the reactant state corresponds to the
bath in its undisplaced, vacuum state, whereas at the potential minimum of the product state each bath mode is displaced by
an amount depending on its frequency and the strength of its coupling to the TLS $\sqrt{J(\omega)}/\omega$. The presence of the reorganization energy in $H_{S}$ ensures that these two minima are degenerate in energy and thus
detailed balance will ensure an equal forward and backward rate.

Turning on the coupling $\epsilon$ between the two levels leads to an avoided crossing in the two energy surfaces in an adiabatic representation of the vibronic tunneling system, leading to two potential wells. In such a semi-classical (Born-Oppeheimer) picture, we see that the electron must overcome a kind of effective energy barrier $E_b$ that scales with the total reorganisation energy of the entire environment $\lambda_R$ in order for the
reaction to progress. We thus might well expect to see thermally activated (exponential) behaviour whereby the tunneling rate $\propto \exp(-\beta E_b)$. However, at low temperatures this behaviour should be dramatically quenched and dissipative quantum tunneling should become dominant and strongly dependent on the spectral function of the environment \citep{weiss_quantum_2012}.   

\subsection{Numerical results}

\begin{figure}
  \includegraphics[width=\columnwidth]{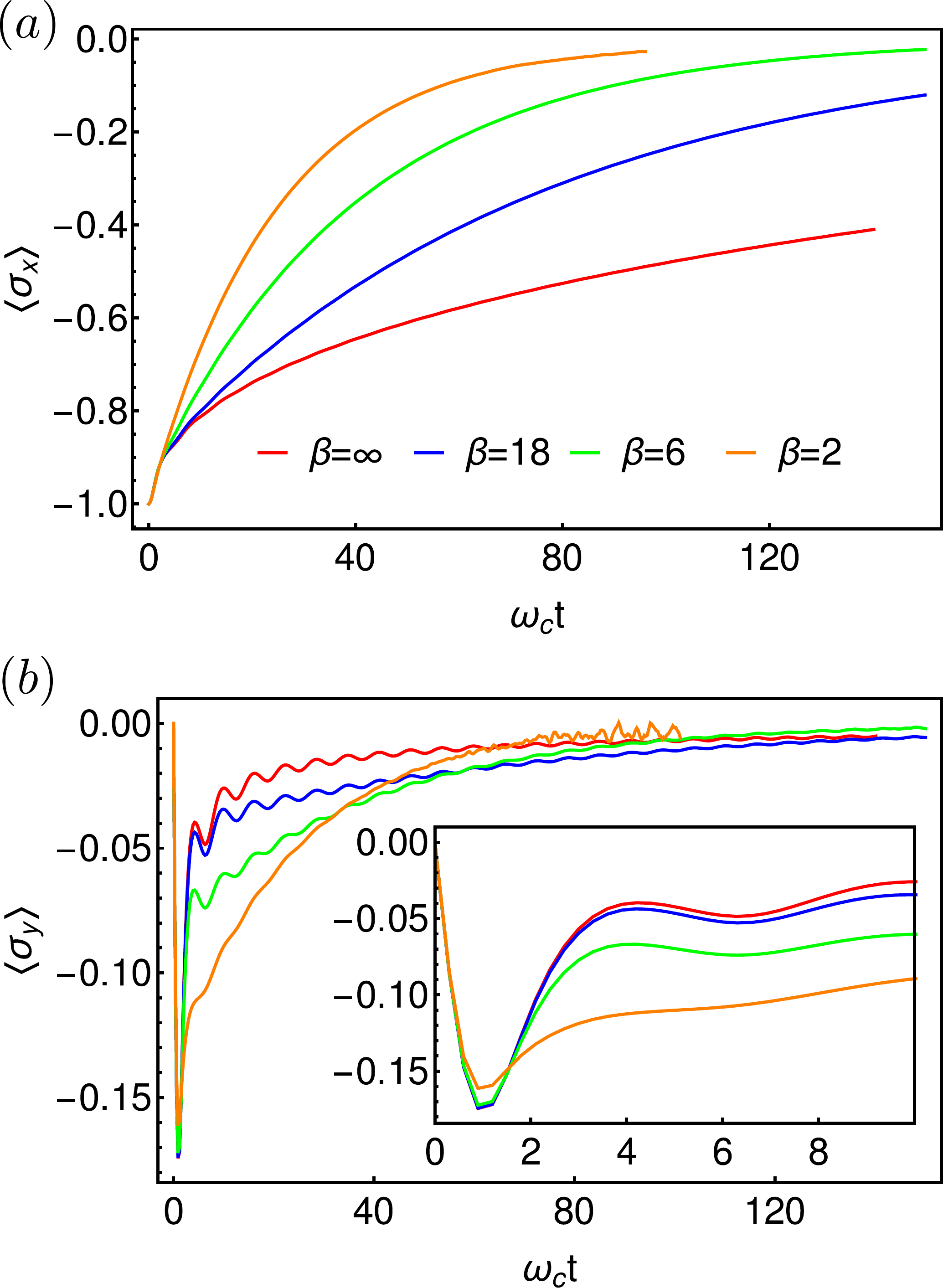}
  \caption{(a) $\langle\sigma_{x}(t)\rangle$ for several temperatures, which represents the progress of the reaction. The
    decay to the steady state is exponential at high temperature. (b) $\langle \sigma_{y}(t)\rangle$, representing the
    momentum along the reaction coordinate. We encounter some noise beyond about $\omega_{c}t=50$ in the $\beta=2$ data.
    This is as a result of the truncation of the local Hilbert spaces of the bath modes (cf. sec
    \ref{sec:numerical-results}). The inset shows an enlarged view of the initial fast dynamics which appear to be
    broadly independent of temperature.}
  \label{fig:spin}
\end{figure}

For our numerical investigation we take an Ohmic spectral density with $\alpha=0.8$ for which the dynamics are expected to be incoherent at all temperatures, i.e. the energy
surfaces of figure \ref{fig:schem}(b) are well separated and friction is such that there will be no oscillatory tunneling dynamics between reactant and
product. In figure \ref{fig:spin} we present results for this model at several temperatures using the T-TEDOPA mapping and 1TDVP. The
expectation of $\sigma_{x}$ can be taken to be a measure of the progress of the reaction, starting at the value of $-1$
when the system is entirely in the reactant state, and approaching $0$ as the electron tunnels through the barrier and
the populations thermalize. We find that as the temperature is increased the dynamics tend to an exponential decay to
the steady state, whereas non-exponential behavior is observed for lower temperatures. In figure \ref{fig:spin}(b) we
show the expectation of $\sigma_{y}$, which is the conjugate coordinate to the $\sigma_{x}$ and which may thus be
interpreted as a kind of momentum associated with the tunneling. We find that there is a sharp initial spike in $\langle
\sigma_{y}\rangle$ which decays with oscillations which are increasingly damped at higher temperatures. As we might have predicted, these transient dynamics occur on a timescale of $\tau\approx \omega_{c}^{-1}$, which the fastest response time of an environment with an upper cut-off frequency of $\omega_c$. This is approximately the timescale over which the environment will adjust to the sudden presence of the electron, and essentially sets the timescale for the formation of the adiabatic landscape (or, alternatively, for the formation of the dressed polaron states), after which the tunneling dynamics proceed. This period is related to the slippage of initial conditions that is sometimes used to fix issues of density matrix positivity in perturbative Redfield Theory \citep{gaspard1999slippage}, although here the establishment of these conditions is described exactly and in real-time. We also see that the crossover to the tunneling regime happens faster as the temperature increases, meaning that the effective initial conditions - particularly $\langle \sigma_y (t) \rangle$ - are  temperature dependent. 

\begin{figure}
  \includegraphics[width=\columnwidth]{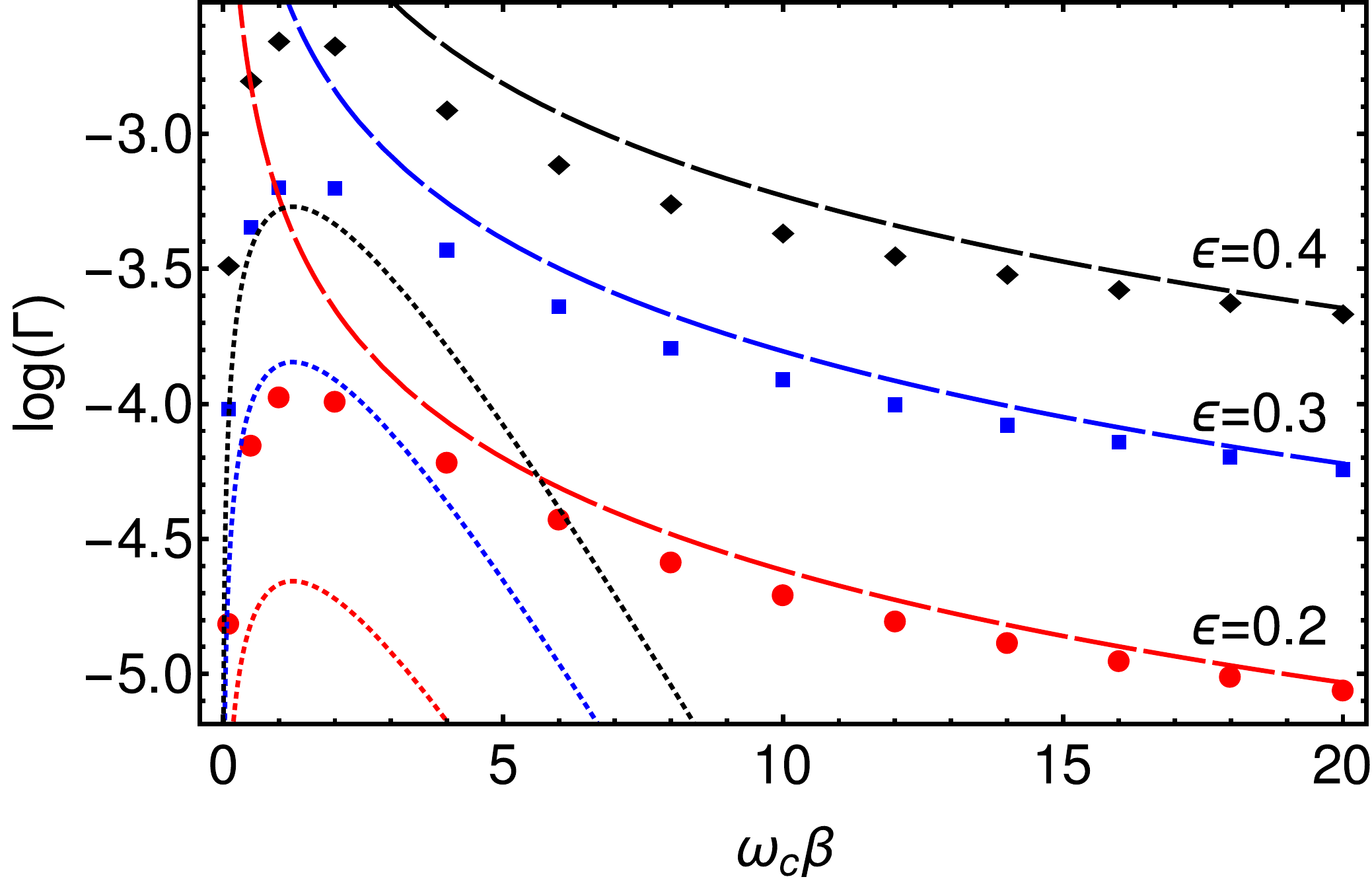}
  \caption{Log plot of the rates, $\Gamma$, extracted from $\langle \sigma_{x}(t)\rangle$ for $\epsilon=0.2$ (Red), $\epsilon=0.3$ (Blue) and
    $\epsilon=0.4$ (Red) as a function of $\beta$. (Dashed lines) High temperature ($T \gg \omega_{c}$), classical, limit of Golden
    Rule formula. (Dotted lines) Low temperature ($T \ll \omega_{c}$), quantum, limit of Golden Rule formula.}
  \label{fig:rates}
\end{figure}

We extract approximate reaction rates from the TLS dynamics by fitting each $\langle\sigma_{x}(t)\rangle$ to an
exponential decay $-e^{-\Gamma t}$ on timescales $t>\tau$. We thus obtain the rates $\Gamma(\epsilon, \beta)$ for the various values of $\beta$
and $\epsilon$ simulated. The values of $\epsilon$ were chosen to be small compared to the characteristic vibrational
frequency of the bath, $\epsilon \ll \omega_{c}$ and to the reorganisation energy, $\epsilon \ll \lambda_{R}$ and thus lie
in the non-adiabatic regime which is the relevant regime for electron transfer. One may then perform a perturbative
expansion in $\epsilon$, otherwise known as the `Golden Rule' approach which, for an Ohmic bath, yields the following
formulas for the high and low temperature limits corresponding respectively to the classical and quantum regimes \citep{weiss_quantum_2012}.
\begin{equation}
  \label{eq:gr}
  \Gamma(\beta) = 
  \begin{cases}
    \frac{\sqrt{\pi}}{4\sqrt{\alpha}}\epsilon^{2}(\frac{\pi}{\beta\omega_{c} })^{2\alpha-1}, \beta\omega_{c} \gg 1\\
    \frac{e^{2}}{4}\sqrt{\frac{\pi\beta\omega_{c} }{2\alpha}}\exp({-\frac{\alpha\beta\omega_c}{2}}), \beta\omega_{c} \ll 1
  \end{cases}
.
\end{equation}
The golden rule result is based on second-order perturbation in the tunneling coupling  $\epsilon$, but it is exact to all orders in the system-environment coupling $\alpha$. Additionally, the Ohmic form of the spectral function generates a non-trivial power-law dependence of the tunneling rate on the temperature for $ \beta\omega_{c} \gg 1$ in which the rate may either decrease or increase as the temperature is lowered, depending on the value of $\alpha$.  We plot these formulas along with the numerically evaluated rates in figure \ref{fig:rates}. There is a good agreement
in the high and low temperature limits between the Golden Rule expressions and the T-TEDOPA results, and one clearly sees that the
temperature dependence of the rate is non-monotonic with a transition from
power law growth (quantum, $2\alpha-1>0$) to power-law decay (classical, $\propto \sqrt{\beta}$) as the temperature increases from $T=0$. We note that for the parameters we present here, the intermediate regime where thermally activated behaviour is predicted $\beta\omega_c \sim1$ is not observed for the Ohmic environment, and one essentially switches from tunneling limited by the effect of friction on the attempt frequency to the low-temperature polaronic tunneling of Eqs. \ref{eq:gr}.  

\section{Conclusion}
\label{sec:con}

\noindent
In this article we have shown how the combination of the Tamasceli's remarkable T-TEDOPA mapping and non-perturbative variational Tensor-Network dynamics can be applied to chemical and photophysical systems under laboratory conditions. Through numerical experiments we have carefully investigated how the T-TEDOPA mapping allows the effects of finite temperatures to be obtained efficiently without any need for costly sampling of the thermal environment state, or the explicit use of density matrices. However, analysis of these environmental dynamics reveals how incorporating finite temperatures can lead to more expensive simulations, due to the filling-up of the chain modes and the longer chains that are needed to prevent recurrence dynamics. Yet, we believe that this
method, and others like it, based on the exact quantum many-body treatment of vibrational modes \citep{somoza2019dissipation}, could present an attractive
complementary approach to the Multi-Layer Multi-Configurational Time-Dependent Hartree Method (MLMCTDH) commonly used in
chemical dynamics. One possible direction for this would be to consider a problem in
which a (discretized) potential surface for a reaction is contained within the system Hamiltonian, while the environment bath provides the
nuclear thermal and quantum fluctuations that ultimately determine both real-time kinetics and thermodynamical yields for the process, as is currently captured in methods such as Ring Polymer Molecular Dynamics \citep{craig2004quantum}. Furthermore, the Tensor-Network structures are not limited to the simple chain
geometries we consider here but can in fact adopt a tree structure, thus enabling the treatment of complex coupling to
multiple independent baths \citep{schroder_tensor_2019}. Such trees tensor networks have recently been interfaced with \textit{ab initio} methods to explore ultra-fast photophysics of real molecules and their pump-probe spectra \citep{schnedermann2019molecular}, but such efforts have so far been limited to zero temperature. 
Finally, the cooperative, antagonistic or sequential actions of different types of environments, i.e. light and vibrations \citep{wertnik2018optimizing}, or even the creation of new excitations, such as polaritons \citep{herrera2020molecular,memmi2017strong,del2018tensor}, could play a key role in sophisticated new materials for energy transduction, catalysis or regulation (feedback) of reactions, and T-TDEPODA-based tensor networks are currently being used to explore these developing areas.  

\bibliography{bibliography2}

\begin{thebibliography}{52}%
\makeatletter
\providecommand \@ifxundefined [1]{%
 \@ifx{#1\undefined}
}%
\providecommand \@ifnum [1]{%
 \ifnum #1\expandafter \@firstoftwo
 \else \expandafter \@secondoftwo
 \fi
}%
\providecommand \@ifx [1]{%
 \ifx #1\expandafter \@firstoftwo
 \else \expandafter \@secondoftwo
 \fi
}%
\providecommand \natexlab [1]{#1}%
\providecommand \enquote  [1]{``#1''}%
\providecommand \bibnamefont  [1]{#1}%
\providecommand \bibfnamefont [1]{#1}%
\providecommand \citenamefont [1]{#1}%
\providecommand \href@noop [0]{\@secondoftwo}%
\providecommand \href [0]{\begingroup \@sanitize@url \@href}%
\providecommand \@href[1]{\@@startlink{#1}\@@href}%
\providecommand \@@href[1]{\endgroup#1\@@endlink}%
\providecommand \@sanitize@url [0]{\catcode `\\12\catcode `\$12\catcode
  `\&12\catcode `\#12\catcode `\^12\catcode `\_12\catcode `\%12\relax}%
\providecommand \@@startlink[1]{}%
\providecommand \@@endlink[0]{}%
\providecommand \url  [0]{\begingroup\@sanitize@url \@url }%
\providecommand \@url [1]{\endgroup\@href {#1}{\urlprefix }}%
\providecommand \urlprefix  [0]{URL }%
\providecommand \Eprint [0]{\href }%
\providecommand \doibase [0]{http://dx.doi.org/}%
\providecommand \selectlanguage [0]{\@gobble}%
\providecommand \bibinfo  [0]{\@secondoftwo}%
\providecommand \bibfield  [0]{\@secondoftwo}%
\providecommand \translation [1]{[#1]}%
\providecommand \BibitemOpen [0]{}%
\providecommand \bibitemStop [0]{}%
\providecommand \bibitemNoStop [0]{.\EOS\space}%
\providecommand \EOS [0]{\spacefactor3000\relax}%
\providecommand \BibitemShut  [1]{\csname bibitem#1\endcsname}%
\let\auto@bib@innerbib\@empty
\bibitem [{\citenamefont {Miller}\ \emph {et~al.}(1983)\citenamefont {Miller},
  \citenamefont {Schwartz},\ and\ \citenamefont {Tromp}}]{miller_quantum_1983}%
  \BibitemOpen
  \bibfield  {author} {\bibinfo {author} {\bibfnamefont {W.~H.}\ \bibnamefont
  {Miller}}, \bibinfo {author} {\bibfnamefont {S.~D.}\ \bibnamefont
  {Schwartz}}, \ and\ \bibinfo {author} {\bibfnamefont {J.~W.}\ \bibnamefont
  {Tromp}},\ }\href {\doibase 10.1063/1.445581} {\bibfield  {journal} {\bibinfo
   {journal} {The Journal of Chemical Physics}\ }\textbf {\bibinfo {volume}
  {79}},\ \bibinfo {pages} {4889} (\bibinfo {year} {1983})},\ \bibinfo {note}
  {publisher: American Institute of Physics}\BibitemShut {NoStop}%
\bibitem [{\citenamefont {Devault}(1980)}]{devault1980quantum}%
  \BibitemOpen
  \bibfield  {author} {\bibinfo {author} {\bibfnamefont {D.}~\bibnamefont
  {Devault}},\ }\href@noop {} {\bibfield  {journal} {\bibinfo  {journal}
  {Quarterly reviews of biophysics}\ }\textbf {\bibinfo {volume} {13}},\
  \bibinfo {pages} {387} (\bibinfo {year} {1980})}\BibitemShut {NoStop}%
\bibitem [{\citenamefont {May}\ and\ \citenamefont
  {K{\"u}hn}(2008)}]{may2008charge}%
  \BibitemOpen
  \bibfield  {author} {\bibinfo {author} {\bibfnamefont {V.}~\bibnamefont
  {May}}\ and\ \bibinfo {author} {\bibfnamefont {O.}~\bibnamefont {K{\"u}hn}},\
  }\href@noop {} {\emph {\bibinfo {title} {Charge and energy transfer dynamics
  in molecular systems}}}\ (\bibinfo  {publisher} {John Wiley \& Sons},\
  \bibinfo {year} {2008})\BibitemShut {NoStop}%
\bibitem [{\citenamefont {Dubi}\ and\ \citenamefont
  {Dia Ventra}(2011)}]{dubi_colloquium_2011}%
  \BibitemOpen
  \bibfield  {author} {\bibinfo {author} {\bibfnamefont {Y.}~\bibnamefont
  {Dubi}}\ and\ \bibinfo {author} {\bibfnamefont {M.}~\bibnamefont
  {Dia Ventra}},\ }\href {\doibase 10.1103/RevModPhys.83.131} {\bibfield
  {journal} {\bibinfo  {journal} {Reviews of Modern Physics}\ }\textbf
  {\bibinfo {volume} {83}},\ \bibinfo {pages} {131} (\bibinfo {year}
  {2011})}\BibitemShut {NoStop}%
\bibitem [{\citenamefont {Benenti}\ \emph {et~al.}(2017)\citenamefont
  {Benenti}, \citenamefont {Casati}, \citenamefont {Saito},\ and\ \citenamefont
  {Whitney}}]{benenti_fundamental_2017}%
  \BibitemOpen
  \bibfield  {author} {\bibinfo {author} {\bibfnamefont {G.}~\bibnamefont
  {Benenti}}, \bibinfo {author} {\bibfnamefont {G.}~\bibnamefont {Casati}},
  \bibinfo {author} {\bibfnamefont {K.}~\bibnamefont {Saito}}, \ and\ \bibinfo
  {author} {\bibfnamefont {R.}~\bibnamefont {Whitney}},\ }\href {\doibase
  10.1016/j.physrep.2017.05.008} {\bibfield  {journal} {\bibinfo  {journal}
  {Physics Reports}\ }\bibinfo {series} {Fundamental aspects of steady-state
  conversion of heat to work at the nanoscale},\ \textbf {\bibinfo {volume}
  {694}},\ \bibinfo {pages} {1} (\bibinfo {year} {2017})}\BibitemShut {NoStop}%
\bibitem [{\citenamefont {Breuer}\ \emph {et~al.}(2002)\citenamefont {Breuer},
  \citenamefont {Petruccione} \emph {et~al.}}]{breuer2002theory}%
  \BibitemOpen
  \bibfield  {author} {\bibinfo {author} {\bibfnamefont {H.-P.}\ \bibnamefont
  {Breuer}}, \bibinfo {author} {\bibfnamefont {F.}~\bibnamefont {Petruccione}},
   \emph {et~al.},\ }\href@noop {} {\emph {\bibinfo {title} {The theory of open
  quantum systems}}}\ (\bibinfo  {publisher} {Oxford University Press on
  Demand},\ \bibinfo {year} {2002})\BibitemShut {NoStop}%
\bibitem [{\citenamefont {Weiss}(2012)}]{weiss_quantum_2012}%
  \BibitemOpen
  \bibfield  {author} {\bibinfo {author} {\bibfnamefont {U.}~\bibnamefont
  {Weiss}},\ }\href {\doibase 10.1142/8334} {\emph {\bibinfo {title} {Quantum
  {Dissipative} {Systems}}}},\ \bibinfo {edition} {4th}\ ed.\ (\bibinfo
  {publisher} {WORLD SCIENTIFIC},\ \bibinfo {year} {2012})\BibitemShut
  {NoStop}%
\bibitem [{\citenamefont {Wang}\ and\ \citenamefont
  {Shao}(2019)}]{wang2019quantum}%
  \BibitemOpen
  \bibfield  {author} {\bibinfo {author} {\bibfnamefont {H.}~\bibnamefont
  {Wang}}\ and\ \bibinfo {author} {\bibfnamefont {J.}~\bibnamefont {Shao}},\
  }\href@noop {} {\bibfield  {journal} {\bibinfo  {journal} {The Journal of
  Physical Chemistry A}\ }\textbf {\bibinfo {volume} {123}},\ \bibinfo {pages}
  {1882} (\bibinfo {year} {2019})}\BibitemShut {NoStop}%
\bibitem [{\citenamefont {Lubich}(2015)}]{lubich_time_2015-1}%
  \BibitemOpen
  \bibfield  {author} {\bibinfo {author} {\bibfnamefont {C.}~\bibnamefont
  {Lubich}},\ }\href {\doibase 10.1093/amrx/abv006} {\bibfield  {journal}
  {\bibinfo  {journal} {Applied Mathematics Research eXpress}\ }\textbf
  {\bibinfo {volume} {2015}},\ \bibinfo {pages} {311} (\bibinfo {year}
  {2015})}\BibitemShut {NoStop}%
\bibitem [{\citenamefont {Orus}(2014)}]{orus_practical_2014}%
  \BibitemOpen
  \bibfield  {author} {\bibinfo {author} {\bibfnamefont {R.}~\bibnamefont
  {Orus}},\ }\href {\doibase 10.1016/j.aop.2014.06.013} {\bibfield  {journal}
  {\bibinfo  {journal} {Annals of Physics}\ }\textbf {\bibinfo {volume}
  {349}},\ \bibinfo {pages} {117} (\bibinfo {year} {2014})},\ \bibinfo {note}
  {arXiv: 1306.2164}\BibitemShut {NoStop}%
\bibitem [{\citenamefont {Prior}\ \emph {et~al.}(2010)\citenamefont {Prior},
  \citenamefont {Chin}, \citenamefont {Huelga},\ and\ \citenamefont
  {Plenio}}]{prior2010efficient}%
  \BibitemOpen
  \bibfield  {author} {\bibinfo {author} {\bibfnamefont {J.}~\bibnamefont
  {Prior}}, \bibinfo {author} {\bibfnamefont {A.~W.}\ \bibnamefont {Chin}},
  \bibinfo {author} {\bibfnamefont {S.~F.}\ \bibnamefont {Huelga}}, \ and\
  \bibinfo {author} {\bibfnamefont {M.~B.}\ \bibnamefont {Plenio}},\
  }\href@noop {} {\bibfield  {journal} {\bibinfo  {journal} {Physical review
  letters}\ }\textbf {\bibinfo {volume} {105}},\ \bibinfo {pages} {050404}
  (\bibinfo {year} {2010})}\BibitemShut {NoStop}%
\bibitem [{\citenamefont {Prior}\ \emph {et~al.}(2013)\citenamefont {Prior},
  \citenamefont {de~Vega}, \citenamefont {Chin}, \citenamefont {Huelga},\ and\
  \citenamefont {Plenio}}]{prior_quantum_2013}%
  \BibitemOpen
  \bibfield  {author} {\bibinfo {author} {\bibfnamefont {J.}~\bibnamefont
  {Prior}}, \bibinfo {author} {\bibfnamefont {I.}~\bibnamefont {de~Vega}},
  \bibinfo {author} {\bibfnamefont {A.~W.}\ \bibnamefont {Chin}}, \bibinfo
  {author} {\bibfnamefont {S.~F.}\ \bibnamefont {Huelga}}, \ and\ \bibinfo
  {author} {\bibfnamefont {M.~B.}\ \bibnamefont {Plenio}},\ }\href {\doibase
  10.1103/PhysRevA.87.013428} {\bibfield  {journal} {\bibinfo  {journal}
  {Physical Review A}\ }\textbf {\bibinfo {volume} {87}},\ \bibinfo {pages}
  {013428} (\bibinfo {year} {2013})},\ \bibinfo {note} {arXiv:
  1205.2897}\BibitemShut {NoStop}%
\bibitem [{\citenamefont {Chin}\ \emph {et~al.}(2013)\citenamefont {Chin},
  \citenamefont {Prior}, \citenamefont {Rosenbach}, \citenamefont
  {Caycedo-Soler}, \citenamefont {Huelga},\ and\ \citenamefont
  {Plenio}}]{chin2013role}%
  \BibitemOpen
  \bibfield  {author} {\bibinfo {author} {\bibfnamefont {A.}~\bibnamefont
  {Chin}}, \bibinfo {author} {\bibfnamefont {J.}~\bibnamefont {Prior}},
  \bibinfo {author} {\bibfnamefont {R.}~\bibnamefont {Rosenbach}}, \bibinfo
  {author} {\bibfnamefont {F.}~\bibnamefont {Caycedo-Soler}}, \bibinfo {author}
  {\bibfnamefont {S.~F.}\ \bibnamefont {Huelga}}, \ and\ \bibinfo {author}
  {\bibfnamefont {M.~B.}\ \bibnamefont {Plenio}},\ }\href@noop {} {\bibfield
  {journal} {\bibinfo  {journal} {Nature Physics}\ }\textbf {\bibinfo {volume}
  {9}},\ \bibinfo {pages} {113} (\bibinfo {year} {2013})}\BibitemShut {NoStop}%
\bibitem [{\citenamefont {Xie}\ \emph {et~al.}(2019)\citenamefont {Xie},
  \citenamefont {Liu}, \citenamefont {Yao}, \citenamefont {Schollwöck},
  \citenamefont {Liu},\ and\ \citenamefont {Ma}}]{xie_time-dependent_2019}%
  \BibitemOpen
  \bibfield  {author} {\bibinfo {author} {\bibfnamefont {X.}~\bibnamefont
  {Xie}}, \bibinfo {author} {\bibfnamefont {Y.}~\bibnamefont {Liu}}, \bibinfo
  {author} {\bibfnamefont {Y.}~\bibnamefont {Yao}}, \bibinfo {author}
  {\bibfnamefont {U.}~\bibnamefont {Schollwöck}}, \bibinfo {author}
  {\bibfnamefont {C.}~\bibnamefont {Liu}}, \ and\ \bibinfo {author}
  {\bibfnamefont {H.}~\bibnamefont {Ma}},\ }\href {\doibase 10.1063/1.5125945}
  {\bibfield  {journal} {\bibinfo  {journal} {The Journal of Chemical Physics}\
  }\textbf {\bibinfo {volume} {151}},\ \bibinfo {pages} {224101} (\bibinfo
  {year} {2019})}\BibitemShut {NoStop}%
\bibitem [{\citenamefont {Alvertis}\ \emph {et~al.}()\citenamefont {Alvertis},
  \citenamefont {Schröder},\ and\ \citenamefont
  {Chin}}]{alvertis_non-equilibrium_2019}%
  \BibitemOpen
  \bibfield  {author} {\bibinfo {author} {\bibfnamefont {A.~M.}\ \bibnamefont
  {Alvertis}}, \bibinfo {author} {\bibfnamefont {F.~A. Y.~N.}\ \bibnamefont
  {Schröder}}, \ and\ \bibinfo {author} {\bibfnamefont {A.~W.}\ \bibnamefont
  {Chin}},\ }\href {\doibase 10.1063/1.5115239} {\bibfield  {journal} {\bibinfo
   {journal} {The Journal of Chemical Physics}\ }\textbf {\bibinfo {volume}
  {151}},\ 10.1063/1.5115239}\BibitemShut {NoStop}%
\bibitem [{\citenamefont {Schröder}\ \emph {et~al.}(2019)\citenamefont
  {Schröder}, \citenamefont {Turban}, \citenamefont {Musser}, \citenamefont
  {Hine},\ and\ \citenamefont {Chin}}]{schroder_tensor_2019}%
  \BibitemOpen
  \bibfield  {author} {\bibinfo {author} {\bibfnamefont {F.~A. Y.~N.}\
  \bibnamefont {Schröder}}, \bibinfo {author} {\bibfnamefont {D.~H.~P.}\
  \bibnamefont {Turban}}, \bibinfo {author} {\bibfnamefont {A.~J.}\
  \bibnamefont {Musser}}, \bibinfo {author} {\bibfnamefont {N.~D.~M.}\
  \bibnamefont {Hine}}, \ and\ \bibinfo {author} {\bibfnamefont {A.~W.}\
  \bibnamefont {Chin}},\ }\href {\doibase 10.1038/s41467-019-09039-7}
  {\bibfield  {journal} {\bibinfo  {journal} {Nature Communications}\ }\textbf
  {\bibinfo {volume} {10}},\ \bibinfo {pages} {1062} (\bibinfo {year}
  {2019})}\BibitemShut {NoStop}%
\bibitem [{\citenamefont {Tamascelli}\ \emph {et~al.}(2019)\citenamefont
  {Tamascelli}, \citenamefont {Smirne}, \citenamefont {Lim}, \citenamefont
  {Huelga},\ and\ \citenamefont {Plenio}}]{tamascelli_efficient_2019}%
  \BibitemOpen
  \bibfield  {author} {\bibinfo {author} {\bibfnamefont {D.}~\bibnamefont
  {Tamascelli}}, \bibinfo {author} {\bibfnamefont {A.}~\bibnamefont {Smirne}},
  \bibinfo {author} {\bibfnamefont {J.}~\bibnamefont {Lim}}, \bibinfo {author}
  {\bibfnamefont {S.~F.}\ \bibnamefont {Huelga}}, \ and\ \bibinfo {author}
  {\bibfnamefont {M.~B.}\ \bibnamefont {Plenio}},\ }\href {\doibase
  10.1103/PhysRevLett.123.090402} {\bibfield  {journal} {\bibinfo  {journal}
  {Physical Review Letters}\ }\textbf {\bibinfo {volume} {123}},\ \bibinfo
  {pages} {090402} (\bibinfo {year} {2019})},\ \bibinfo {note} {arXiv:
  1811.12418}\BibitemShut {NoStop}%
\bibitem [{\citenamefont {Wilhelm}\ \emph {et~al.}(2004)\citenamefont
  {Wilhelm}, \citenamefont {Kleff},\ and\ \citenamefont
  {Von~Delft}}]{wilhelm2004spin}%
  \BibitemOpen
  \bibfield  {author} {\bibinfo {author} {\bibfnamefont {F.}~\bibnamefont
  {Wilhelm}}, \bibinfo {author} {\bibfnamefont {S.}~\bibnamefont {Kleff}}, \
  and\ \bibinfo {author} {\bibfnamefont {J.}~\bibnamefont {Von~Delft}},\
  }\href@noop {} {\bibfield  {journal} {\bibinfo  {journal} {Chemical physics}\
  }\textbf {\bibinfo {volume} {296}},\ \bibinfo {pages} {345} (\bibinfo {year}
  {2004})}\BibitemShut {NoStop}%
\bibitem [{\citenamefont {Schulze}\ and\ \citenamefont
  {Kuhn}(2015)}]{schulze2015explicit}%
  \BibitemOpen
  \bibfield  {author} {\bibinfo {author} {\bibfnamefont {J.}~\bibnamefont
  {Schulze}}\ and\ \bibinfo {author} {\bibfnamefont {O.}~\bibnamefont {Kuhn}},\
  }\href@noop {} {\bibfield  {journal} {\bibinfo  {journal} {The Journal of
  Physical Chemistry B}\ }\textbf {\bibinfo {volume} {119}},\ \bibinfo {pages}
  {6211} (\bibinfo {year} {2015})}\BibitemShut {NoStop}%
\bibitem [{\citenamefont {Mendive-Tapia}\ \emph {et~al.}(2018)\citenamefont
  {Mendive-Tapia}, \citenamefont {Mangaud}, \citenamefont {Firmino},
  \citenamefont {de~la Lande}, \citenamefont {Desouter-Lecomte}, \citenamefont
  {Meyer},\ and\ \citenamefont {Gatti}}]{mendive2018multidimensional}%
  \BibitemOpen
  \bibfield  {author} {\bibinfo {author} {\bibfnamefont {D.}~\bibnamefont
  {Mendive-Tapia}}, \bibinfo {author} {\bibfnamefont {E.}~\bibnamefont
  {Mangaud}}, \bibinfo {author} {\bibfnamefont {T.}~\bibnamefont {Firmino}},
  \bibinfo {author} {\bibfnamefont {A.}~\bibnamefont {de~la Lande}}, \bibinfo
  {author} {\bibfnamefont {M.}~\bibnamefont {Desouter-Lecomte}}, \bibinfo
  {author} {\bibfnamefont {H.-D.}\ \bibnamefont {Meyer}}, \ and\ \bibinfo
  {author} {\bibfnamefont {F.}~\bibnamefont {Gatti}},\ }\href@noop {}
  {\bibfield  {journal} {\bibinfo  {journal} {The Journal of Physical Chemistry
  B}\ }\textbf {\bibinfo {volume} {122}},\ \bibinfo {pages} {126} (\bibinfo
  {year} {2018})}\BibitemShut {NoStop}%
\bibitem [{\citenamefont {Mukamel}(1995)}]{mukamel1995principles}%
  \BibitemOpen
  \bibfield  {author} {\bibinfo {author} {\bibfnamefont {S.}~\bibnamefont
  {Mukamel}},\ }\href@noop {} {\emph {\bibinfo {title} {Principles of nonlinear
  optical spectroscopy}}},\ Vol.~\bibinfo {volume} {6}\ (\bibinfo  {publisher}
  {Oxford university press New York},\ \bibinfo {year} {1995})\BibitemShut
  {NoStop}%
\bibitem [{\citenamefont {G{\'e}linas}\ \emph {et~al.}(2014)\citenamefont
  {G{\'e}linas}, \citenamefont {Rao}, \citenamefont {Kumar}, \citenamefont
  {Smith}, \citenamefont {Chin}, \citenamefont {Clark}, \citenamefont {van~der
  Poll}, \citenamefont {Bazan},\ and\ \citenamefont
  {Friend}}]{gelinas2014ultra-fast}%
  \BibitemOpen
  \bibfield  {author} {\bibinfo {author} {\bibfnamefont {S.}~\bibnamefont
  {G{\'e}linas}}, \bibinfo {author} {\bibfnamefont {A.}~\bibnamefont {Rao}},
  \bibinfo {author} {\bibfnamefont {A.}~\bibnamefont {Kumar}}, \bibinfo
  {author} {\bibfnamefont {S.~L.}\ \bibnamefont {Smith}}, \bibinfo {author}
  {\bibfnamefont {A.~W.}\ \bibnamefont {Chin}}, \bibinfo {author}
  {\bibfnamefont {J.}~\bibnamefont {Clark}}, \bibinfo {author} {\bibfnamefont
  {T.~S.}\ \bibnamefont {van~der Poll}}, \bibinfo {author} {\bibfnamefont
  {G.~C.}\ \bibnamefont {Bazan}}, \ and\ \bibinfo {author} {\bibfnamefont
  {R.~H.}\ \bibnamefont {Friend}},\ }\href@noop {} {\bibfield  {journal}
  {\bibinfo  {journal} {Science}\ }\textbf {\bibinfo {volume} {343}},\ \bibinfo
  {pages} {512} (\bibinfo {year} {2014})}\BibitemShut {NoStop}%
\bibitem [{\citenamefont {Smith}\ and\ \citenamefont
  {Chin}(2015)}]{smith2015phonon}%
  \BibitemOpen
  \bibfield  {author} {\bibinfo {author} {\bibfnamefont {S.~L.}\ \bibnamefont
  {Smith}}\ and\ \bibinfo {author} {\bibfnamefont {A.~W.}\ \bibnamefont
  {Chin}},\ }\href@noop {} {\bibfield  {journal} {\bibinfo  {journal} {Physical
  Review B}\ }\textbf {\bibinfo {volume} {91}},\ \bibinfo {pages} {201302}
  (\bibinfo {year} {2015})}\BibitemShut {NoStop}%
\bibitem [{\citenamefont {Alvermann}\ and\ \citenamefont
  {Fehske}(2009)}]{alvermann2009sparse}%
  \BibitemOpen
  \bibfield  {author} {\bibinfo {author} {\bibfnamefont {A.}~\bibnamefont
  {Alvermann}}\ and\ \bibinfo {author} {\bibfnamefont {H.}~\bibnamefont
  {Fehske}},\ }\href@noop {} {\bibfield  {journal} {\bibinfo  {journal}
  {Physical review letters}\ }\textbf {\bibinfo {volume} {102}},\ \bibinfo
  {pages} {150601} (\bibinfo {year} {2009})}\BibitemShut {NoStop}%
\bibitem [{\citenamefont {Binder}\ and\ \citenamefont
  {Burghardt}(2019)}]{binder2019first}%
  \BibitemOpen
  \bibfield  {author} {\bibinfo {author} {\bibfnamefont {R.}~\bibnamefont
  {Binder}}\ and\ \bibinfo {author} {\bibfnamefont {I.}~\bibnamefont
  {Burghardt}},\ }\href@noop {} {\bibfield  {journal} {\bibinfo  {journal}
  {Faraday Discussions}\ }\textbf {\bibinfo {volume} {221}},\ \bibinfo {pages}
  {406} (\bibinfo {year} {2019})}\BibitemShut {NoStop}%
\bibitem [{\citenamefont {Tamascelli}\ \emph {et~al.}(2018)\citenamefont
  {Tamascelli}, \citenamefont {Smirne}, \citenamefont {Huelga},\ and\
  \citenamefont {Plenio}}]{tamascelli2018nonperturbative}%
  \BibitemOpen
  \bibfield  {author} {\bibinfo {author} {\bibfnamefont {D.}~\bibnamefont
  {Tamascelli}}, \bibinfo {author} {\bibfnamefont {A.}~\bibnamefont {Smirne}},
  \bibinfo {author} {\bibfnamefont {S.~F.}\ \bibnamefont {Huelga}}, \ and\
  \bibinfo {author} {\bibfnamefont {M.~B.}\ \bibnamefont {Plenio}},\
  }\href@noop {} {\bibfield  {journal} {\bibinfo  {journal} {Physical review
  letters}\ }\textbf {\bibinfo {volume} {120}},\ \bibinfo {pages} {030402}
  (\bibinfo {year} {2018})}\BibitemShut {NoStop}%
\bibitem [{\citenamefont {Musser}\ \emph {et~al.}(2015)\citenamefont {Musser},
  \citenamefont {Liebel}, \citenamefont {Schnedermann}, \citenamefont {Wende},
  \citenamefont {Kehoe}, \citenamefont {Rao},\ and\ \citenamefont
  {Kukura}}]{musser2015evidence}%
  \BibitemOpen
  \bibfield  {author} {\bibinfo {author} {\bibfnamefont {A.~J.}\ \bibnamefont
  {Musser}}, \bibinfo {author} {\bibfnamefont {M.}~\bibnamefont {Liebel}},
  \bibinfo {author} {\bibfnamefont {C.}~\bibnamefont {Schnedermann}}, \bibinfo
  {author} {\bibfnamefont {T.}~\bibnamefont {Wende}}, \bibinfo {author}
  {\bibfnamefont {T.~B.}\ \bibnamefont {Kehoe}}, \bibinfo {author}
  {\bibfnamefont {A.}~\bibnamefont {Rao}}, \ and\ \bibinfo {author}
  {\bibfnamefont {P.}~\bibnamefont {Kukura}},\ }\href@noop {} {\bibfield
  {journal} {\bibinfo  {journal} {Nature Physics}\ }\textbf {\bibinfo {volume}
  {11}},\ \bibinfo {pages} {352} (\bibinfo {year} {2015})}\BibitemShut
  {NoStop}%
\bibitem [{\citenamefont {Schnedermann}\ \emph {et~al.}(2016)\citenamefont
  {Schnedermann}, \citenamefont {Lim}, \citenamefont {Wende}, \citenamefont
  {Duarte}, \citenamefont {Ni}, \citenamefont {Gu}, \citenamefont {Sadhanala},
  \citenamefont {Rao},\ and\ \citenamefont {Kukura}}]{schnedermann2016sub}%
  \BibitemOpen
  \bibfield  {author} {\bibinfo {author} {\bibfnamefont {C.}~\bibnamefont
  {Schnedermann}}, \bibinfo {author} {\bibfnamefont {J.~M.}\ \bibnamefont
  {Lim}}, \bibinfo {author} {\bibfnamefont {T.}~\bibnamefont {Wende}}, \bibinfo
  {author} {\bibfnamefont {A.~S.}\ \bibnamefont {Duarte}}, \bibinfo {author}
  {\bibfnamefont {L.}~\bibnamefont {Ni}}, \bibinfo {author} {\bibfnamefont
  {Q.}~\bibnamefont {Gu}}, \bibinfo {author} {\bibfnamefont {A.}~\bibnamefont
  {Sadhanala}}, \bibinfo {author} {\bibfnamefont {A.}~\bibnamefont {Rao}}, \
  and\ \bibinfo {author} {\bibfnamefont {P.}~\bibnamefont {Kukura}},\
  }\href@noop {} {\bibfield  {journal} {\bibinfo  {journal} {The journal of
  physical chemistry letters}\ }\textbf {\bibinfo {volume} {7}},\ \bibinfo
  {pages} {4854} (\bibinfo {year} {2016})}\BibitemShut {NoStop}%
\bibitem [{\citenamefont {Schnedermann}\ \emph {et~al.}(2019)\citenamefont
  {Schnedermann}, \citenamefont {Alvertis}, \citenamefont {Wende},
  \citenamefont {Lukman}, \citenamefont {Feng}, \citenamefont {Schr{\"o}der},
  \citenamefont {Turban}, \citenamefont {Wu}, \citenamefont {Hine},
  \citenamefont {Greenham} \emph {et~al.}}]{schnedermann2019molecular}%
  \BibitemOpen
  \bibfield  {author} {\bibinfo {author} {\bibfnamefont {C.}~\bibnamefont
  {Schnedermann}}, \bibinfo {author} {\bibfnamefont {A.~M.}\ \bibnamefont
  {Alvertis}}, \bibinfo {author} {\bibfnamefont {T.}~\bibnamefont {Wende}},
  \bibinfo {author} {\bibfnamefont {S.}~\bibnamefont {Lukman}}, \bibinfo
  {author} {\bibfnamefont {J.}~\bibnamefont {Feng}}, \bibinfo {author}
  {\bibfnamefont {F.~A.}\ \bibnamefont {Schr{\"o}der}}, \bibinfo {author}
  {\bibfnamefont {D.~H.}\ \bibnamefont {Turban}}, \bibinfo {author}
  {\bibfnamefont {J.}~\bibnamefont {Wu}}, \bibinfo {author} {\bibfnamefont
  {N.~D.}\ \bibnamefont {Hine}}, \bibinfo {author} {\bibfnamefont {N.~C.}\
  \bibnamefont {Greenham}},  \emph {et~al.},\ }\href@noop {} {\bibfield
  {journal} {\bibinfo  {journal} {Nature communications}\ }\textbf {\bibinfo
  {volume} {10}},\ \bibinfo {pages} {1} (\bibinfo {year} {2019})}\BibitemShut
  {NoStop}%
\bibitem [{\citenamefont {de~Vega}\ and\ \citenamefont
  {Bañuls}()}]{de_vega_thermofield-based_2015}%
  \BibitemOpen
  \bibfield  {author} {\bibinfo {author} {\bibfnamefont {I.}~\bibnamefont
  {de~Vega}}\ and\ \bibinfo {author} {\bibfnamefont {M.-C.}\ \bibnamefont
  {Bañuls}},\ }\href {\doibase 10.1103/PhysRevA.92.052116} {\bibfield
  {journal} {\bibinfo  {journal} {Physical Review A}\ }\textbf {\bibinfo
  {volume} {92}},\ 10.1103/PhysRevA.92.052116}\BibitemShut {NoStop}%
\bibitem [{\citenamefont {Chin}\ \emph {et~al.}(2010)\citenamefont {Chin},
  \citenamefont {Rivas}, \citenamefont {Huelga},\ and\ \citenamefont
  {Plenio}}]{chin2010exact}%
  \BibitemOpen
  \bibfield  {author} {\bibinfo {author} {\bibfnamefont {A.~W.}\ \bibnamefont
  {Chin}}, \bibinfo {author} {\bibfnamefont {{\'A}.}~\bibnamefont {Rivas}},
  \bibinfo {author} {\bibfnamefont {S.~F.}\ \bibnamefont {Huelga}}, \ and\
  \bibinfo {author} {\bibfnamefont {M.~B.}\ \bibnamefont {Plenio}},\
  }\href@noop {} {\bibfield  {journal} {\bibinfo  {journal} {Journal of
  Mathematical Physics}\ }\textbf {\bibinfo {volume} {51}},\ \bibinfo {pages}
  {092109} (\bibinfo {year} {2010})}\BibitemShut {NoStop}%
\bibitem [{\citenamefont {Schollwöck}()}]{schollwock_density-matrix_2011}%
  \BibitemOpen
  \bibfield  {author} {\bibinfo {author} {\bibfnamefont {U.}~\bibnamefont
  {Schollwöck}},\ }\href {\doibase 10.1016/j.aop.2010.09.012} {\bibfield
  {journal} {\bibinfo  {journal} {Annals of Physics}\ }\textbf {\bibinfo
  {volume} {326}},\ 10.1016/j.aop.2010.09.012}\BibitemShut {NoStop}%
\bibitem [{\citenamefont {Lubich}\ \emph {et~al.}(2015)\citenamefont {Lubich},
  \citenamefont {Oseledets},\ and\ \citenamefont
  {Vandereycken}}]{lubich_time_2015}%
  \BibitemOpen
  \bibfield  {author} {\bibinfo {author} {\bibfnamefont {C.}~\bibnamefont
  {Lubich}}, \bibinfo {author} {\bibfnamefont {I.}~\bibnamefont {Oseledets}}, \
  and\ \bibinfo {author} {\bibfnamefont {B.}~\bibnamefont {Vandereycken}},\
  }\href {\doibase 10.1137/140976546} {\bibfield  {journal} {\bibinfo
  {journal} {SIAM Journal on Numerical Analysis}\ }\textbf {\bibinfo {volume}
  {53}},\ \bibinfo {pages} {917} (\bibinfo {year} {2015})},\ \bibinfo {note}
  {arXiv: 1407.2042}\BibitemShut {NoStop}%
\bibitem [{\citenamefont {Paeckel}\ \emph {et~al.}()\citenamefont {Paeckel},
  \citenamefont {Köhler}, \citenamefont {Swoboda}, \citenamefont {Manmana},
  \citenamefont {Schollwöck},\ and\ \citenamefont
  {Hubig}}]{paeckel_time-evolution_2019}%
  \BibitemOpen
  \bibfield  {author} {\bibinfo {author} {\bibfnamefont {S.}~\bibnamefont
  {Paeckel}}, \bibinfo {author} {\bibfnamefont {T.}~\bibnamefont {Köhler}},
  \bibinfo {author} {\bibfnamefont {A.}~\bibnamefont {Swoboda}}, \bibinfo
  {author} {\bibfnamefont {S.~R.}\ \bibnamefont {Manmana}}, \bibinfo {author}
  {\bibfnamefont {U.}~\bibnamefont {Schollwöck}}, \ and\ \bibinfo {author}
  {\bibfnamefont {C.}~\bibnamefont {Hubig}},\ }\href {\doibase
  10.1016/j.aop.2019.167998} {\bibfield  {journal} {\bibinfo  {journal} {Annals
  of Physics}\ }\textbf {\bibinfo {volume} {411}},\
  10.1016/j.aop.2019.167998}\BibitemShut {NoStop}%
\bibitem [{\citenamefont {Haegeman}\ \emph {et~al.}(2016)\citenamefont
  {Haegeman}, \citenamefont {Lubich}, \citenamefont {Oseledets}, \citenamefont
  {Vandereycken},\ and\ \citenamefont {Verstraete}}]{haegeman_unifying_2016}%
  \BibitemOpen
  \bibfield  {author} {\bibinfo {author} {\bibfnamefont {J.}~\bibnamefont
  {Haegeman}}, \bibinfo {author} {\bibfnamefont {C.}~\bibnamefont {Lubich}},
  \bibinfo {author} {\bibfnamefont {I.}~\bibnamefont {Oseledets}}, \bibinfo
  {author} {\bibfnamefont {B.}~\bibnamefont {Vandereycken}}, \ and\ \bibinfo
  {author} {\bibfnamefont {F.}~\bibnamefont {Verstraete}},\ }\href {\doibase
  10.1103/PhysRevB.94.165116} {\bibfield  {journal} {\bibinfo  {journal}
  {Physical Review B}\ }\textbf {\bibinfo {volume} {94}},\ \bibinfo {pages}
  {165116} (\bibinfo {year} {2016})},\ \bibinfo {note} {arXiv:
  1408.5056}\BibitemShut {NoStop}%
\bibitem [{\citenamefont {Woods}\ \emph {et~al.}()\citenamefont {Woods},
  \citenamefont {Cramer},\ and\ \citenamefont
  {Plenio}}]{woods_simulating_2015}%
  \BibitemOpen
  \bibfield  {author} {\bibinfo {author} {\bibfnamefont {M.}~\bibnamefont
  {Woods}}, \bibinfo {author} {\bibfnamefont {M.}~\bibnamefont {Cramer}}, \
  and\ \bibinfo {author} {\bibfnamefont {M.}~\bibnamefont {Plenio}},\ }\href
  {\doibase 10.1103/PhysRevLett.115.130401} {\bibfield  {journal} {\bibinfo
  {journal} {Physical Review Letters}\ }\textbf {\bibinfo {volume} {115}},\
  10.1103/PhysRevLett.115.130401}\BibitemShut {NoStop}%
\bibitem [{\citenamefont {Schr{\"o}der}\ and\ \citenamefont
  {Chin}(2016)}]{schroder2016simulating}%
  \BibitemOpen
  \bibfield  {author} {\bibinfo {author} {\bibfnamefont {F.~A.}\ \bibnamefont
  {Schr{\"o}der}}\ and\ \bibinfo {author} {\bibfnamefont {A.~W.}\ \bibnamefont
  {Chin}},\ }\href@noop {} {\bibfield  {journal} {\bibinfo  {journal} {Physical
  Review B}\ }\textbf {\bibinfo {volume} {93}},\ \bibinfo {pages} {075105}
  (\bibinfo {year} {2016})}\BibitemShut {NoStop}%
\bibitem [{\citenamefont {Mahan}(2000)}]{mahan_many-particle_2000}%
  \BibitemOpen
  \bibfield  {author} {\bibinfo {author} {\bibfnamefont {G.~D.}\ \bibnamefont
  {Mahan}},\ }\href {\doibase 10.1007/978-1-4757-5714-9} {{\selectlanguage
  {en}\emph {\bibinfo {title} {Many-{Particle} {Physics}}}}}\ (\bibinfo
  {publisher} {Springer US},\ \bibinfo {address} {Boston, MA},\ \bibinfo {year}
  {2000})\BibitemShut {NoStop}%
\bibitem [{\citenamefont {De~Vega}\ and\ \citenamefont
  {Alonso}(2017)}]{de2017dynamics}%
  \BibitemOpen
  \bibfield  {author} {\bibinfo {author} {\bibfnamefont {I.}~\bibnamefont
  {De~Vega}}\ and\ \bibinfo {author} {\bibfnamefont {D.}~\bibnamefont
  {Alonso}},\ }\href@noop {} {\bibfield  {journal} {\bibinfo  {journal}
  {Reviews of Modern Physics}\ }\textbf {\bibinfo {volume} {89}},\ \bibinfo
  {pages} {015001} (\bibinfo {year} {2017})}\BibitemShut {NoStop}%
\bibitem [{\citenamefont {Silbey}\ and\ \citenamefont
  {Harris}(1984)}]{silbey_variational_1984}%
  \BibitemOpen
  \bibfield  {author} {\bibinfo {author} {\bibfnamefont {R.}~\bibnamefont
  {Silbey}}\ and\ \bibinfo {author} {\bibfnamefont {R.~A.}\ \bibnamefont
  {Harris}},\ }\href {\doibase 10.1063/1.447055} {\bibfield  {journal}
  {\bibinfo  {journal} {The Journal of Chemical Physics}\ }\textbf {\bibinfo
  {volume} {80}},\ \bibinfo {pages} {2615} (\bibinfo {year}
  {1984})}\BibitemShut {NoStop}%
\bibitem [{\citenamefont {Guo}\ \emph {et~al.}(2012)\citenamefont {Guo},
  \citenamefont {Weichselbaum}, \citenamefont {von Delft},\ and\ \citenamefont
  {Vojta}}]{guo2012critical}%
  \BibitemOpen
  \bibfield  {author} {\bibinfo {author} {\bibfnamefont {C.}~\bibnamefont
  {Guo}}, \bibinfo {author} {\bibfnamefont {A.}~\bibnamefont {Weichselbaum}},
  \bibinfo {author} {\bibfnamefont {J.}~\bibnamefont {von Delft}}, \ and\
  \bibinfo {author} {\bibfnamefont {M.}~\bibnamefont {Vojta}},\ }\href@noop {}
  {\bibfield  {journal} {\bibinfo  {journal} {Physical review letters}\
  }\textbf {\bibinfo {volume} {108}},\ \bibinfo {pages} {160401} (\bibinfo
  {year} {2012})}\BibitemShut {NoStop}%
\bibitem [{\citenamefont {Marcus}(1993)}]{marcus1993electron}%
  \BibitemOpen
  \bibfield  {author} {\bibinfo {author} {\bibfnamefont {R.~A.}\ \bibnamefont
  {Marcus}},\ }\href@noop {} {\bibfield  {journal} {\bibinfo  {journal}
  {Reviews of Modern Physics}\ }\textbf {\bibinfo {volume} {65}},\ \bibinfo
  {pages} {599} (\bibinfo {year} {1993})}\BibitemShut {NoStop}%
\bibitem [{\citenamefont {Zuehlsdorff}\ \emph {et~al.}(2019)\citenamefont
  {Zuehlsdorff}, \citenamefont {Montoya-Castillo}, \citenamefont {Napoli},
  \citenamefont {Markland},\ and\ \citenamefont
  {Isborn}}]{zuehlsdorff_optical_2019}%
  \BibitemOpen
  \bibfield  {author} {\bibinfo {author} {\bibfnamefont {T.~J.}\ \bibnamefont
  {Zuehlsdorff}}, \bibinfo {author} {\bibfnamefont {A.}~\bibnamefont
  {Montoya-Castillo}}, \bibinfo {author} {\bibfnamefont {J.~A.}\ \bibnamefont
  {Napoli}}, \bibinfo {author} {\bibfnamefont {T.~E.}\ \bibnamefont
  {Markland}}, \ and\ \bibinfo {author} {\bibfnamefont {C.~M.}\ \bibnamefont
  {Isborn}},\ }\href {\doibase 10.1063/1.5114818} {\bibfield  {journal}
  {\bibinfo  {journal} {The Journal of Chemical Physics}\ }\textbf {\bibinfo
  {volume} {151}},\ \bibinfo {pages} {074111} (\bibinfo {year}
  {2019})}\BibitemShut {NoStop}%
\bibitem [{\citenamefont {Womick}\ \emph {et~al.}(2011)\citenamefont {Womick},
  \citenamefont {Liu},\ and\ \citenamefont {Moran}}]{womick2011exciton}%
  \BibitemOpen
  \bibfield  {author} {\bibinfo {author} {\bibfnamefont {J.~M.}\ \bibnamefont
  {Womick}}, \bibinfo {author} {\bibfnamefont {H.}~\bibnamefont {Liu}}, \ and\
  \bibinfo {author} {\bibfnamefont {A.~M.}\ \bibnamefont {Moran}},\ }\href@noop
  {} {\bibfield  {journal} {\bibinfo  {journal} {The Journal of Physical
  Chemistry A}\ }\textbf {\bibinfo {volume} {115}},\ \bibinfo {pages} {2471}
  (\bibinfo {year} {2011})}\BibitemShut {NoStop}%
\bibitem [{\citenamefont {Kolli}\ \emph {et~al.}(2012)\citenamefont {Kolli},
  \citenamefont {O'Reilly}, \citenamefont {Scholes},\ and\ \citenamefont
  {Olaya-Castro}}]{kolli2012fundamental}%
  \BibitemOpen
  \bibfield  {author} {\bibinfo {author} {\bibfnamefont {A.}~\bibnamefont
  {Kolli}}, \bibinfo {author} {\bibfnamefont {E.~J.}\ \bibnamefont {O'Reilly}},
  \bibinfo {author} {\bibfnamefont {G.~D.}\ \bibnamefont {Scholes}}, \ and\
  \bibinfo {author} {\bibfnamefont {A.}~\bibnamefont {Olaya-Castro}},\
  }\href@noop {} {\bibfield  {journal} {\bibinfo  {journal} {The Journal of
  chemical physics}\ }\textbf {\bibinfo {volume} {137}},\ \bibinfo {pages}
  {174109} (\bibinfo {year} {2012})}\BibitemShut {NoStop}%
\bibitem [{\citenamefont {Gaspard}\ and\ \citenamefont
  {Nagaoka}(1999)}]{gaspard1999slippage}%
  \BibitemOpen
  \bibfield  {author} {\bibinfo {author} {\bibfnamefont {P.}~\bibnamefont
  {Gaspard}}\ and\ \bibinfo {author} {\bibfnamefont {M.}~\bibnamefont
  {Nagaoka}},\ }\href@noop {} {\bibfield  {journal} {\bibinfo  {journal} {The
  Journal of chemical physics}\ }\textbf {\bibinfo {volume} {111}},\ \bibinfo
  {pages} {5668} (\bibinfo {year} {1999})}\BibitemShut {NoStop}%
\bibitem [{\citenamefont {Somoza}\ \emph {et~al.}(2019)\citenamefont {Somoza},
  \citenamefont {Marty}, \citenamefont {Lim}, \citenamefont {Huelga},\ and\
  \citenamefont {Plenio}}]{somoza2019dissipation}%
  \BibitemOpen
  \bibfield  {author} {\bibinfo {author} {\bibfnamefont {A.~D.}\ \bibnamefont
  {Somoza}}, \bibinfo {author} {\bibfnamefont {O.}~\bibnamefont {Marty}},
  \bibinfo {author} {\bibfnamefont {J.}~\bibnamefont {Lim}}, \bibinfo {author}
  {\bibfnamefont {S.~F.}\ \bibnamefont {Huelga}}, \ and\ \bibinfo {author}
  {\bibfnamefont {M.~B.}\ \bibnamefont {Plenio}},\ }\href@noop {} {\bibfield
  {journal} {\bibinfo  {journal} {Physical Review Letters}\ }\textbf {\bibinfo
  {volume} {123}},\ \bibinfo {pages} {100502} (\bibinfo {year}
  {2019})}\BibitemShut {NoStop}%
\bibitem [{\citenamefont {Craig}\ and\ \citenamefont
  {Manolopoulos}(2004)}]{craig2004quantum}%
  \BibitemOpen
  \bibfield  {author} {\bibinfo {author} {\bibfnamefont {I.~R.}\ \bibnamefont
  {Craig}}\ and\ \bibinfo {author} {\bibfnamefont {D.~E.}\ \bibnamefont
  {Manolopoulos}},\ }\href@noop {} {\bibfield  {journal} {\bibinfo  {journal}
  {The Journal of chemical physics}\ }\textbf {\bibinfo {volume} {121}},\
  \bibinfo {pages} {3368} (\bibinfo {year} {2004})}\BibitemShut {NoStop}%
\bibitem [{\citenamefont {Wertnik}\ \emph {et~al.}(2018)\citenamefont
  {Wertnik}, \citenamefont {Chin}, \citenamefont {Nori},\ and\ \citenamefont
  {Lambert}}]{wertnik2018optimizing}%
  \BibitemOpen
  \bibfield  {author} {\bibinfo {author} {\bibfnamefont {M.}~\bibnamefont
  {Wertnik}}, \bibinfo {author} {\bibfnamefont {A.}~\bibnamefont {Chin}},
  \bibinfo {author} {\bibfnamefont {F.}~\bibnamefont {Nori}}, \ and\ \bibinfo
  {author} {\bibfnamefont {N.}~\bibnamefont {Lambert}},\ }\href@noop {}
  {\bibfield  {journal} {\bibinfo  {journal} {The Journal of chemical physics}\
  }\textbf {\bibinfo {volume} {149}},\ \bibinfo {pages} {084112} (\bibinfo
  {year} {2018})}\BibitemShut {NoStop}%
\bibitem [{\citenamefont {Herrera}\ and\ \citenamefont
  {Owrutsky}(2020)}]{herrera2020molecular}%
  \BibitemOpen
  \bibfield  {author} {\bibinfo {author} {\bibfnamefont {F.}~\bibnamefont
  {Herrera}}\ and\ \bibinfo {author} {\bibfnamefont {J.}~\bibnamefont
  {Owrutsky}},\ }\href@noop {} {\bibfield  {journal} {\bibinfo  {journal} {The
  Journal of Chemical Physics}\ }\textbf {\bibinfo {volume} {152}},\ \bibinfo
  {pages} {100902} (\bibinfo {year} {2020})}\BibitemShut {NoStop}%
\bibitem [{\citenamefont {Memmi}\ \emph {et~al.}(2017)\citenamefont {Memmi},
  \citenamefont {Benson}, \citenamefont {Sadofev},\ and\ \citenamefont
  {Kalusniak}}]{memmi2017strong}%
  \BibitemOpen
  \bibfield  {author} {\bibinfo {author} {\bibfnamefont {H.}~\bibnamefont
  {Memmi}}, \bibinfo {author} {\bibfnamefont {O.}~\bibnamefont {Benson}},
  \bibinfo {author} {\bibfnamefont {S.}~\bibnamefont {Sadofev}}, \ and\
  \bibinfo {author} {\bibfnamefont {S.}~\bibnamefont {Kalusniak}},\ }\href@noop
  {} {\bibfield  {journal} {\bibinfo  {journal} {Physical review letters}\
  }\textbf {\bibinfo {volume} {118}},\ \bibinfo {pages} {126802} (\bibinfo
  {year} {2017})}\BibitemShut {NoStop}%
\bibitem [{\citenamefont {Del~Pino}\ \emph {et~al.}(2018)\citenamefont
  {Del~Pino}, \citenamefont {Schr{\"o}der}, \citenamefont {Chin}, \citenamefont
  {Feist},\ and\ \citenamefont {Garcia-Vidal}}]{del2018tensor}%
  \BibitemOpen
  \bibfield  {author} {\bibinfo {author} {\bibfnamefont {J.}~\bibnamefont
  {Del~Pino}}, \bibinfo {author} {\bibfnamefont {F.~A.}\ \bibnamefont
  {Schr{\"o}der}}, \bibinfo {author} {\bibfnamefont {A.~W.}\ \bibnamefont
  {Chin}}, \bibinfo {author} {\bibfnamefont {J.}~\bibnamefont {Feist}}, \ and\
  \bibinfo {author} {\bibfnamefont {F.~J.}\ \bibnamefont {Garcia-Vidal}},\
  }\href@noop {} {\bibfield  {journal} {\bibinfo  {journal} {Physical Review
  B}\ }\textbf {\bibinfo {volume} {98}},\ \bibinfo {pages} {165416} (\bibinfo
  {year} {2018})}\BibitemShut {NoStop}%
\end{thebibliography}%


%

\section*{Author Contributions}
AJD implemented the T-TEDOPA mapping in a bespoke 1TDVP code and performed the numerical simulations. AJD and AWC wrote the manuscript. AWC oversaw the project. 

\section*{Code}
The codes used in this work are freely available for reasonable use at  \href{https://github.com/angusdunnett/MPSDynamics}{https://github.com/angusdunnett/MPSDynamics}.  
\section*{Funding}
AJD is supported by the Ecole Doctorale 564 `Physique en Ile-de-France'. AWC is partly supported by ANR project No. 195608/ACCEPT.

\onecolumngrid
\newpage
\section{Appendix}
The chain mapping used in section \ref{sec:t-tedopa} is based on the theory of orthogonal polynomials. A polynomial of degree $n$ is defined by
\begin{equation}
    p_{n}(x)=\sum_{m=0}^{n}a_{m}x^m.
\end{equation}
The space of polynomials of degree $n$ is denoted $\mathbb{P}_{n}$ and is a subset of the space of all polynomials $\mathbb{P}_{n} \subset \mathbb{P}$.
Given a measure $d\mu(x)$ which has finite moments of all orders on some interval $[a,b]$, we may define the inner product of two polynomials 
\begin{equation}
    \langle{p,q}\rangle_{\mu} = \int_{a}^{b}d\mu(x)p(x)q(x).
\end{equation}
This inner product gives rise to a unique set of orthonormal polynomials $\{\tilde{p}_{n} \in \mathbb{P}_{n}, n=0,1,2,...\}$ which all satisfy
\begin{equation}
    \langle\tilde{p}_{n},\tilde{p}_{m}\rangle=\delta_{n,m}.
\end{equation}
This set forms a complete basis for $\mathbb{P}$, and more specifically the set $\{\tilde{p}_{n} \in \mathbb{P}_{n}, n=0,1,2, ... m\}$ is a complete basis for $\bigcup_{r=1}^{m}\mathbb{P}_{r}$.

It is often useful to express the orthonormal polynomials in terms of the orthogonal \emph{monic} polynomials $\pi_{n}(x)$ which are the unnormalized scalar multiples of $\tilde{p}_{n}(x)$ whose leading coefficient is 1 ($a_{n}=1$) 
\begin{equation}
  \tilde{p}_{n}(x)=\frac{\pi_{n}(x)}{||\pi_{n}||}.
\end{equation}

The key property of orthogonal polynomials for the construction of the chain mapping is that they satisfy a three term recurrence relation 
\begin{equation}
\label{eq:three}
    \pi_{k+1}(x) = (x-\alpha_{k})\pi_{k}(x)-\beta_{k}\pi_{k-1}(x),
\end{equation}
where it can be easily shown that 
\begin{equation}
\label{eq:ab}
    \alpha_{k}=\frac{\langle x\pi_{k},\pi_{k}\rangle}{\langle\pi_{k},\pi_{k}\rangle}, \beta_{k}=\frac{\langle\pi_{k},\pi_{k}\rangle}{\langle\pi_{k-1},\pi_{k-1}\rangle}.
\end{equation}
Now that we have defined the orthogonal polynomials we may use them to construct the unitary transformation that will convert the star Hamiltonian of Eq. \ref{eq:Hext} with
\begin{equation}
  \label{eq:ext2}
  H_{I}^{\text{ext}}=A_{S}\otimes\int_{-\infty}^{\infty}d\omega\sqrt{J_{\beta}(\omega)}(a_{\omega}+a_{\omega}^{\dagger}), 
  H_{E}^{\text{ext}}=\int_{-\infty}^{\infty}d\omega\omega a_{\omega}^{\dagger}a_{\omega},
\end{equation}
into the chain Hamiltonian of Eq. \ref{eq:chain}. The transformation is given by 
\begin{equation}
    c_{n}^{(\dagger)}=\int_{-\infty}^{\infty}U_{n}(\omega)a_{\omega}^{(\dagger)},
\end{equation}
where 
\begin{equation}
    U_{n}(\omega) = \sqrt{J_{\beta}(\omega)}\tilde{p}_{n}(\omega) = \sqrt{J_{\beta}(\omega)}\frac{\pi_{n}(\omega)}{||\pi_{n}||},
\end{equation}
and the polynomials $\tilde{p}_{n}(\omega)$ are orthonormal with respect to the measure $d\omega J_{\beta}(\omega)$.
The unitarity of $U_{n}(\omega)$ follows immediately from the orthonormality of the polynomials.

Applying the above transformation to the interaction Hamiltonian we have
\begin{equation}
    H_{I}^{\text{ext}}=A_{S}\otimes\sum_{n=0}^{\infty}\int_{-\infty}^{\infty}d\omega J_{\beta}(\omega)\frac{\pi_{n}(\omega)}{||\pi_{n}||}(c_{n}^{\dagger}+c_{n})
\end{equation}
For the zeroth order monic polynomial we have $\pi_{0}=1$ and so we may insert this into the above expression
\begin{equation}
    H_{I}^{\text{ext}}=A_{S}\otimes\sum_{n=0}^{\infty}\int_{-\infty}^{\infty}d\omega J_{\beta}(\omega)\frac{\pi_{n}(\omega)\pi_{0}}{||\pi_{n}||}(c_{n}^{\dagger}+c_{n}).
\end{equation}
Recognising the inner product in the above expression and making use of the orthogonality of the polynomials we have
\begin{equation}
    H_{I}^{\text{ext}}=A_{S}\otimes\sum_{n=0}^{\infty} ||\pi_{n}||\delta_{n,0}(c_{n}^{\dagger}+c_{n})=
    A_{S}\otimes||\pi_{0}||(c_{0}^{\dagger}+c_{0}),
\end{equation}
and thus, in the new basis, only one mode now couples to the system. 

Now for the environment part of the Hamiltonian we have
\begin{equation}
    H_{E}^{\text{ext}}=\sum_{n,m=0}^{\infty}\int_{-\infty}^{\infty}d\omega J_{\beta}(\omega)\omega \frac{\pi_{n}(\omega)\pi_{m}(\omega)}{||\pi_{n}||||\pi_{m}||}c_{n}^{\dagger}c_{m}.
\end{equation}
Substituting for $\omega\pi_{n}(\omega)$ from the three term recurrence relation of Eq. \ref{eq:three} yields
\begin{equation}
     H_{E}^{\text{ext}}=\sum_{n,m=0}^{\infty}\int_{-\infty}^{\infty}d\omega \frac{J_{\beta}(\omega)}{||\pi_{n}||||\pi_{m}||}\Big[\pi_{n+1}(\omega)+\alpha_{n}\pi_{n}(\omega)+\beta_{n}\pi_{n-1}(\omega)\Big]\pi_{m}(\omega)c_{n}^{\dagger}c_{m}.
\end{equation}
Again, evaluating the inner products we have

\begin{equation}
\label{eq:chain2}
\begin{split}
    H_{E}^{\text{ext}}
    &=\sum_{n,m=0}^{\infty}\frac{1}{||\pi_{n}||}\Big[||\pi_{m}||\delta_{n+1,m}+\alpha_{n}||\pi_{m}||\delta_{n,m}+\beta_{n}||\pi_{m}||\delta_{n-1,m}\Big]c_{n}^{\dagger}c_{m}\\
    &=\sum_{n=0}^{\infty}\sqrt{\beta_{n+1}}c_{n}^{\dagger}c_{n+1} + \alpha_{n}c_{n}^{\dagger}c_{n} + \sqrt{\beta_{n+1}}c_{n}^{\dagger}c_{n-1},
\end{split}
\end{equation}
where in the second line we have used the fact that 
\begin{equation}
    \frac{||\pi_{n+1}||}{||\pi_{n}||}=\sqrt{\beta_{n+1}}.
\end{equation}
We thus arrive at the nearest-neighbour coupling Hamiltonian of Eq. \ref{eq:chain} and are able to identify the chain coefficients as
\begin{equation}
\begin{split}
    &\kappa=||\pi_{0}||, \\
    &\omega_{n+1}=\alpha_{n}, \\
    &t_{n}=\sqrt{\beta_{n}}.
\end{split}
\end{equation}
Note that in Eq. \ref{eq:chain} the chain sites are labeled starting from $n=1$ and not $n=0$ as in Eq. \ref{eq:chain2}. 
All that remains now to calculate the chain coefficients for a particular spectral density $J_{\beta}(\omega)$ is to compute the recurrence coefficients, $\alpha_{n}$ and $\beta_{n}$, and this may done interatively using Eqs \ref{eq:three} and \ref{eq:ab} and numerically evaluating the inner product integrals using a quadrature rule.

\end{document}